\documentclass[a4paper,11pt]{article}
\usepackage{jheppub} 
\usepackage{lineno}
\usepackage[dvipsnames, svgnames, x11names]{xcolor}
\usepackage{amsmath}
\usepackage{amssymb}
\usepackage{graphicx}
\usepackage{float}
\usepackage{color}
\usepackage{braket}
\usepackage{orcidlink}
\graphicspath{{figures}}
\usepackage{multicol}
\usepackage{01_wrap}

\newcommand{\EqBBGKYBody}{%
    &
        \frac{\partial \rP_{(n)}(\{\bph\})}{\partial t} 
        + \sum_{i=1}^n \frac{\bp_i}{p_i^0}\cdot\frac{\partial \rP_{(n)}(\{\bph\})}{\partial \bx_i}
    \\=\;&
        \sum_{\substack{i,j=1\\i < j}}^{n}
        \delta^{(3)}(\bx_i-\bx_j)
        \frac{1}{p_i^0 p_j^0}
        \int_{\bp'_i,\bp'_j}
        \mW_{(\bp_i,\bp_j\to \bp_i',\bp_j')}
    \condbreak{\times}
        \big(\rP_{(n)}(\bp_i', \bp_j', \cdots) - \rP_{(n)}(\bp_i, \bp_j, \cdots)\big)
    \\+\;&
        (N-n)\sum_{i=1}^{n}
        \frac{1}{p_i^0}
        \int_{\bp'_i, \bp'_j, \bp_j}
        \mW_{(\bp_i,\bp_j\to \bp_i',\bp_j')}
    \condbreak{\times}
        \big(\rP_{(n+1)}(\bp_i', \cdots; \bp_j') - \rP_{(n+1)}(\bp_i, \cdots; \bp_j)\big)\,,
}
\DeclareEqCmd{EqBBGKY}{eq:QuantumBBGKY}{\EqBBGKYBody}[eqwrap]

\newcommand{\EqbasisBody}{%
\condamp
    \basis_{n,\ell,m}(p_{\mu})  
\condeq
    e^{-p_{\mu} u^\mu / \Lambda}  
    \left( \frac{p_{\mu} u^\mu}{\Lambda} \right)^\ell  
    Y_{\ell,m}(\theta,\phi)  
    L_n^{(2\ell+2)}\left( \frac{p_{\mu} u^\mu}{\Lambda} \right)  
    \,,  
}
\DeclareEqCmd{Eqbasis}{eq:basis}{\EqbasisBody}[eqwrap]

\newcommand{\EqsBBGKYBody}{%
&
    \frac{\partial}{\partial t} 
    \rP^{i_1 i_2 \cdots i_n}(t, \bx_1, \bx_2, \cdots, \bx_n)  
\condbreak{+}
    \sum_{a=1}^n 
    \boldsymbol{B}_{i_a j_a}
    \cdot\frac{\partial }{\partial \bx_a}
    \rP^{i_1 i_2 \cdots j_a \cdots i_n}(t, \bx_1, \bx_2, \cdots, \bx_n)   
\\=\;& 
    \sum_{\substack{a,b=1\\a < b}}^{n}
    \delta^{(3)}(\bx_a-\bx_b)
    C_{i_a i_b j_a j_b}
\condbreak{\times}
    \rP^{i_1 i_2 \cdots j_a \cdots j_b \cdots i_n}(t, \bx_1, \bx_2, \cdots, \bx_n)   
\\+\;&
    (N-n)\sum_{a=1}^{n}
    A_{i_a j_a j_b}
\condbreak{\times}
    \rP^{i_1 i_2 \cdots j_a \cdots i_n j_b}(t, \bx_1, \bx_2, \cdots, \bx_a , \cdots, \bx_n, \bx_a)\,,
}
\DeclareEqCmd{EqsBBGKY}{eq:spectralBBGKY}{\EqsBBGKYBody}[eqwrap]

\newcommand{\EqinttersotBBody}{%
    \boldsymbol{B}_{ij}
=\;&
    \int_{p_\mu} 
    \basisleft_{i}(p_{\mu})
    \basis_{j}(p_{\mu})
    \frac{\bp}{p^0} 
    \;,
}
\DeclareEqCmd{EqinttersotB}{eq:Free streaming Integral}{\EqinttersotBBody}[eqwrap]

\newcommand{\EqdualbasisBody}{%
    \basisleft_{n,\ell,m}(p_{\mu})  
=\;&
    \frac{n!}{(2\ell+n+2)!}  
    \left( \frac{p_{\mu} u^\mu}{\Lambda} \right)^{\ell+2}  
\condbreak{\times}
    Y_{\ell,m}(\theta,\phi)  
    L_n^{(2\ell+2)}\left( \frac{p_{\mu} u^\mu}{\Lambda} \right)  
    \,.
}
\DeclareEqCmd{Eqdualbasis}{eq:dual_basis}{\EqdualbasisBody}[eqwrap]

\newcommand{\EqorthocondBody}{%
    \int_{p_{\mu}}
    \basis_{n,\ell,m}(p_{\mu}) 
    \basisleft_{n',\ell',m'}(p_{\mu})
    = \delta_{n,n'} \delta_{\ell,\ell'} \delta_{m,m'} \,,  
}
\DeclareEqCmd{Eqorthocond}{eq:orthogonality condition}{\EqorthocondBody}[eqwrap]

\newcommand{\EqinttersotATwotermsBody}{%
    A_{ijk} =\;& A^{\mathrm{gain}}_{ijk} - A^{\mathrm{loss}}_{ijk} \,,
}
\DeclareEqCmd{EqinttersotATwoterms}{eq:A_original}{\EqinttersotATwotermsBody}[eqwrap]

\newcommand{\EqinttersotCTwotermsBody}{%
    C_{ijks} =\;& C^{\mathrm{gain}}_{ijks} - C^{\mathrm{loss}}_{ijks} \,.
}
\DeclareEqCmd{EqinttersotCTwoterms}{eq:C_original}{\EqinttersotCTwotermsBody}[eqwrap]

\newcommand{\EqinttersotAGainBody}{%
    A^{\mathrm{gain}}_{ijk}
=\;&
    \int_{p_{1\mu}}
    \int_{\bp_2,\bp_{3},\bp_{4}}
    \basisleft_{i}(p_{1\mu})
\condbreak{\times}
    \frac{1}{p_1^0}
    \mW_{(\bp_1,\bp_2 \rightarrow \bp_{3},\bp_{4})}
    \basis_j(p_{3\mu})
    \basis_k(p_{4\mu})
    \;,
}
\newcommand{\EqinttersotALossBody}{%
    A^{\mathrm{loss}}_{ijk}
=\;&
    \int_{p_{1\mu}}
    \int_{\bp_2,\bp_{3},\bp_{4}}
    \basisleft_{i}(p_{1\mu})
\condbreak{\times}
    \frac{1}{p_1^0}
    \mW_{(\bp_1,\bp_2 \rightarrow \bp_{3},\bp_{4})}
    \basis_j(p_{1\mu})
    \basis_k(p_{2\mu})
    \;,
}
\newcommand{\EqinttersotCGainBody}{%
    C^{\mathrm{gain}}_{ijks}
=\;&
    \int_{p_{1\mu},p_{2\mu}}
    \int_{\bp_{3},\bp_{4}}
    \basisleft_{i}(p_{1\mu})
    \basisleft_{j}(p_{2\mu})
\condbreak{\times}
    \frac{1}{p_1^0 p_2^0}
    \mW_{(\bp_1,\bp_2 \rightarrow \bp_{3},\bp_{4})}
    \basis_k(p_{3\mu})
    \basis_s(p_{4\mu})\;,
}
\newcommand{\EqinttersotCLossBody}{%
    C^{\mathrm{loss}}_{ijks}
=\;&
    \int_{p_{1\mu},p_{2\mu}}
    \int_{\bp_{3},\bp_{4}}
    \basisleft_{i}(p_{1\mu})
    \basisleft_{j}(p_{2\mu})
\condbreak{\times}
    \frac{1}{p_1^0 p_2^0}
    \mW_{(\bp_1,\bp_2 \rightarrow \bp_{3},\bp_{4})}
    \basis_k(p_{1\mu})
    \basis_s(p_{2\mu})\,.
}
\DeclareEqCmd{EqinttersotAGain}{eq:A_Gain}{\EqinttersotAGainBody}[eqwrap]
\DeclareEqCmd{EqinttersotALoss}{eq:A_Loss}{\EqinttersotALossBody}[eqwrap]
\DeclareEqCmd{EqinttersotCGain}{eq:C_Gain}{\EqinttersotCGainBody}[eqwrap]
\DeclareEqCmd{EqinttersotCLoss}{eq:C_Loss}{\EqinttersotCLossBody}[eqwrap]

\usepackage{paracol} 
\usepackage{soul}
\usepackage{empheq}

\newcommand{\bx}{\boldsymbol{x}} 
\newcommand{\bp}{\boldsymbol{p}} 
\newcommand{\rP}{\mathrm{P}}
\newcommand{\bph}{\boldsymbol{\phi}}
\newcommand{\mW}{\mathcal{W}}
\newcommand{\basis}{\mathcal{P}}
\newcommand{\basisleft}{\mathcal{Q}}
\newcommand{\mass}{m_m}
\allowdisplaybreaks[2]

\title{Decoupling hydrodynamization from thermalization via nonlinear Boltzmann equation}

\author[a]{Xingjian Lu}
\author[a]{Shuzhe Shi}
\affiliation[a]{Department of Physics, Tsinghua University,
Beijing 100084, China.}
\emailAdd{lus21@mails.tsinghua.edu.cn}
\emailAdd{shuzhe-shi@tsinghua.edu.cn}

\abstract{
The early thermalization puzzle arises from the unexpectedly early applicability of hydrodynamics in heavy-ion collisions. While hydrodynamics has traditionally been associated with the onset of local thermal equilibrium, its derivations---whether microscopic or macroscopic---rely instead on linearization around equilibrium. However, the linearization timescale---the time at which a system's evolution begins to follow a linearized equation---has not been systematically investigated. In this work, we employ the spectral nonlinear Boltzmann equation---the lowest-order truncation of the spectral Bogoliubov--Born--Green--Kirkwood--Yvon (BBGKY) hierarchy---to analyze the timescales of linearization and thermalization under three distinct truncation schemes. The first two truncations allow for analytic treatment via recursive spectral equations, while the third requires numerical methods for generic initial conditions. The analysis is performed for a homogeneous, massless system with a constant differential cross section. For this simplified setup, we find a robust separation: the linearization time is consistently about half the thermalization time ($\tau_{\mathrm{lin}}/\tau_{\mathrm{therm}} \approx 1/2$). This separation of timescales suggests an explanation for the early applicability of hydrodynamics and points toward a possible quantitative resolution of the early thermalization puzzle. 
}

\begin{document}
\maketitle
\flushbottom

\section{Introduction}

Relativistic heavy-ion collisions were originally proposed as a means to investigate the nature of the QCD vacuum in the 1970s~\cite{Lee:1974kn}, contrasting the physical vacuum---characterized by spontaneous chiral symmetry breaking and color confinement---with the perturbative QCD vacuum, where asymptotic freedom ensures weakly interacting quarks and gluons. In the physical QCD vacuum, the interaction between quarks and gluons is inherently nonperturbative, as the running coupling constant becomes large in the low-energy regime. This leads to a highly nontrivial vacuum structure, characterized by a nonzero quark condensate signaling spontaneous chiral symmetry breaking, a nonzero gluon condensate contributing to the vacuum energy density~\cite{Shifman:1978bx, Shifman:1978by, Colangelo:2000dp}, and the phenomenon of color confinement. To probe these nonperturbative features, it is natural to contrast them with the perturbative QCD regime~\cite{Lee:1974kn}, realized at sufficiently high energy scales where the running coupling becomes small and perturbative methods are valid. It was precisely for this purpose that relativistic heavy-ion collisions were proposed. At the extreme energy densities achieved in such collisions, the QCD vacuum is expected to undergo a transition in which confinement is lifted and chiral symmetry (in the light-quark sector) is approximately restored, allowing quarks and gluons to emerge as effective quasiparticle degrees of freedom. This deconfined phase is commonly referred to as the quark–gluon plasma (QGP). 

Results from the Relativistic Heavy Ion Collider (RHIC) and the Large Hadron Collider (LHC) established that the QGP is a deconfined, strongly coupled fluid, overturning early weak-coupling expectations. During the 1970s–1990s, it was commonly expected that the dynamical evolution of the QGP---from its initial formation to final hadronization---could be adequately described using perturbative QCD techniques~\cite{Shuryak:2003xe}. However, starting from the 2000s, the experimental programs at RHIC and LHC revealed key features that challenged this view. First, equation-of-state calculations from lattice QCD incorporating a phase transition provided a better description of experimental data than that assuming quasi-free quarks and gluons. On a separate note, simulations based on a purely hadronic gas also lead to results inconsistent with data, corroborating the existence of a deconfined QGP phase. Second, observations of collective flow phenomena---such as radial and elliptic flow---were quantitatively reproduced by hydrodynamic simulations exhibiting nearly ideal fluid behavior~\cite{Shen:2014vra, Schenke:2010rr, Karpenko:2013wva, vanderSchee:2013pia, Pang:2018zzo, Du:2019obx}. The most striking result to emerge from these hydrodynamic studies is the extraction of an exceptionally small shear viscosity to entropy density ratio, indicating that the QGP behaves as a strongly coupled fluid---a conclusion that stands in sharp contrast to the weakly interacting gas originally envisioned.

Building upon these insights, the so-called \textit{early thermalization puzzle} arises. This puzzle refers to the apparent tension between the early applicability of hydrodynamic and the expected thermalization timescales of a system initially far from equilibrium~\cite{Heinz:2002un, Shuryak:2003xe}. The hydrodynamic framework presupposes that each fluid cell is near local thermal equilibrium. Phenomenologically, successful descriptions of experimental data using hydrodynamics require its validity to begin extremely early, around $0.6\!-\!1~\mathrm{fm}/c$ after the collision~\cite{Shen:2014vra, Nijs:2020roc, Nijs:2020ors}. This timescale is significantly shorter than conventional estimations based on mean free paths and differential scattering cross-sections in a far-from-equilibrium system.

The \textit{early thermalization puzzle} concerns how far-from-equilibrium initial states in non-Abelian gauge theories achieve thermalization. To date, two main categories of models have been proposed to describe the initial conditions of heavy-ion collisions: incoherent models and coherent models. In incoherent descriptions, the system is viewed as an accumulation of independent parton scatterings that give rise to minijets~\cite{Wang:1991hta, Gyulassy:1994ew}. These minijets then undergo further interactions, ultimately generating a deconfined, thermalized partonic medium. Minijet production is described using perturbative QCD, while the equilibration of the system is modelled on kinetic theory (usually, the relativistic Boltzmann equation) with perturbative QCD parton-parton scattering rates~\cite{Zhou:2024ysb}. Coherent pictures, on the other hand, emphasize the role of collective field configurations. Immediately after the collision, strong longitudinal color-electric fields may be established, which subsequently convert into real partons through the Schwinger mechanism~\cite{Schwinger:1951nm, Casher:1978wy, 1983AnPhy.145..340A, Gyulassy:1985oqt, Matsui:1986xp}. Once liberated, these partons approach thermal equilibrium through interactions encoded in kinetic theory (usually, the relativistic Boltzmann equation), supplemented by the influence of the residual background field. Another well-studied realization of a coherent initial state is the Color Glass Condensate (CGC)~\cite{Gelis:2010nm, Kovner:1995ts, Kovner:1995ja, Krasnitz:1999wc, Krasnitz:2000gz, Schenke:2012wb}, which describes the dense, classical gluon fields carried by the colliding nuclei at small partonic momentum fractions.

Addressing this puzzle---namely, how a strongly coupled system initially far from equilibrium can thermalize so rapidly---requires, in principle, a treatment based on the full Bogoliubov--Born--Green--Kirkwood--Yvon (BBGKY) hierarchy~\cite{bogoliubov1946kineticEnglish, bogoliubov1946kineticRussian, bogoliubov1947kinetic,1946JChPh..14..180K,1947JChPh..15...72K,1946RSPSA.188...10B, yvon1935theorie}. The BBGKY hierarchy offers a time-reversal-symmetric framework to characterize the dynamical evolution of many-body systems far from equilibrium. However, due to the considerable analytic and computational challenges in solving the complete BBGKY equations, a new approach called \textit{spectral BBGKY} has recently been proposed by the authors in Ref.~\cite{Lu:2025yry}. This method expands the reduced distribution functions in a complete orthogonal basis in momentum-space, thereby simplifying the numerical evolution of the BBGKY hierarchy. The basis functions are specifically chosen to efficiently capture thermalization processes in multiparticle systems. In this work, as a preliminary investigation, we begin by studying the lowest-order truncation of this framework---referred to as the \textit{spectral Boltzmann equation}---to analyze the conditions for the emergence of hydrodynamic behavior.

Over the past two decades, explanations of early thermalization have followed two main approaches: modifications of hydrodynamic modeling, in order to capture more early-time dynamics, have progressively relaxed the assumption of near-instantaneous local equilibrium; and kinetic theory, which seeks to derive this behavior from first principles. From the hydrodynamic perspective, the most prominent attempt to address early thermalization is anisotropic hydrodynamics~\cite{Martinez:2010sc, Martinez:2012tu, Ryblewski:2012rr, Bazow:2013ifa, Bazow:2015cha, Tinti:2015xwa, Molnar:2016vvu, McNelis:2021zji, McNelis:2018jho}, which extends the near-isotropy assumption of conventional formulations such as Navier--Stokes and allows for substantial momentum-space anisotropies, reflected by the pressure anisotropy. This relaxation of isotropy provides a more realistic description of the highly anisotropic early stage. Nevertheless, current derivations of anisotropic hydrodynamic equations lack a complete microscopic foundation, particularly a rigorous link to kinetic theory with full collision kernels. From the kinetic-theory perspective, previous studies have primarily focused on two other directions. First, inelastic processes such as $gg \leftrightarrow ggg$ have been shown to play a decisive role in driving the system toward equilibrium~\cite{Xu:2004mz}; their inclusion enables the calculated elliptic flow to agree with RHIC data~\cite{Xu:2007jv, Xu:2008av} and yields shear-viscosity–to–entropy-density ratios $\eta/s$~\cite{Xu:2007jv, Xu:2007ns} as small as the AdS/CFT lower bound~\cite{Kovtun:2004de}. Second, the potential influence of Bose enhancement on thermalization has been explored~\cite{Blaizot:2011xf, Blaizot:2013lga, Xu:2014ega}, though its quantitative impact remains uncertain despite being considered potentially significant.

\vspace{5mm}
Within kinetic theory, a key unresolved issue---central to the present work---concerns the timescale required for a system to reach the \textit{linearization} regime, namely the stage where its evolution can be faithfully described by a linearized approximation around equilibrium. Whereas previous studies have often equated the applicability of hydrodynamics with the onset of local thermalization in each fluid cell, we emphasize a crucial misconception: the foundation of hydrodynamics requires only that the system dynamics follow equations of motion truncated to linear order in the non-equilibrium correction, not that the system be fully thermalized. Despite its importance, this linearization time has not yet been quantitatively determined. Determining this timescale is key to probing the microscopic origin of hydrodynamics and to testing its applicability from first principles. Within kinetic theory, the present work concerns the timescale required for a system to reach the \textit{linearization} regime, namely the stage where its evolution can be faithfully described by a linearized approximation around equilibrium.

In this work, we employ the spectral Boltzmann equation, the lowest-order truncation of the sBBGKY hierarchy~\cite{Lu:2025yry}, to efficiently evolve the distribution function with full, nonlinear collision kernels. We quantitatively determine the linearization timescale $\tau_{\mathrm{lin}}$ and the thermalization timescale $\tau_{\mathrm{therm}}$.  We observe a robust relation, $\tau_{\mathrm{lin}} \approx \tau_{\mathrm{therm}}/2$, suggesting a quantitative resolution of the early-thermalization puzzle.

This paper is organized as follows. The theoretical foundations of the sBBGKY framework relevant to our analysis, along with the assumptions adopted in this study, are presented in Section~\ref{sec:Method}. The linearization and thermalization behavior of a many-body system governed by the nonlinear Boltzmann equation is examined in Section~\ref{sec:Result}, using three distinct spectral truncations. Two of these allow a fully analytical treatment, offering direct insight into nonlinear dynamics, while the third is investigated numerically, enabling an extension to more general initial conditions and providing a unified view of the linearization and thermalization processes. Conclusions are summarized in Section~\ref{sec:Summary}.
\section{The Spectrum Method for the BBGKY hierarchy}\label{sec:Method}

To capture thermalization and linearization processes in a far-from-equilibrium many-body system, we must go beyond the linear approximation and incorporate genuinely \textit{nonlinear} effects. Linearized kinetic descriptions are computationally convenient but reliable only in the vicinity of equilibrium. By contrast, existing nonlinear approaches often become prohibitively expensive and typically evaluate collision integrals using Monte Carlo estimators, thereby limiting controllable accuracy. 

Therefore, we employ the sBBGKY hierarchy recently developed in Ref.~\cite{Lu:2025yry}. The BBGKY hierarchy is a set of coupled equations governing the evolution of reduced distribution functions in many-particle systems. These reduced distributions represent the marginal distributions of the full phase-space distribution function. In the sBBGKY approach, each (reduced) distribution is expanded in a complete set of orthogonal basis functions in momentum space. This transforms the conventional BBGKY hierarchy into sBBGKY, which is analytically equivalent to the original formulation but significantly more amenable to numerical implementation. First, in the spectral expansion, the collision integrals reduce to time-independent, precomputable collision tensors fixed by the chosen basis, enabling closed-form evaluation of collision integrals that are otherwise notoriously difficult to compute analytically. Second, the orthonormal, complete basis is carefully chosen so that the first few modes are sufficient to capture the system's thermalization dynamics, thereby substantially reducing computational cost relative to conventional momentum-space discretization methods. 

In this work, we utilize the lowest-order truncation of this framework---referred to as the \textit{spectral Boltzmann equation}---as a minimal model that incorporates nonlinearity while remaining computationally tractable. Our goal is to analyze the conditions for the emergence of hydrodynamic behavior. In Section~\ref{sec:Method:overview_sBBGKY}, we briefly review the sBBGKY hierarchy for completeness, and readers interested in detailed derivations are referred to Section~II of~\cite{Lu:2025yry}. In Section~\ref{sec:Method:spectral_Boltzmann}, we introduce the lowest-order truncation of the sBBGKY hierarchy---namely, the spectral Boltzmann equation---which serves as the primary equation analyzed in this study. Finally, in Section~\ref{sec:Method:notation}, we summarize the notation and terminology used throughout the paper to ensure clarity.

\subsection{The BBGKY Hierarchy}\label{sec:Method:overview_sBBGKY}

The sBBGKY framework is rooted in the conventional BBGKY hierarchy, which reads
\EqBBGKY
where the momentum-integration measure is defined as $\int_{\bp}\equiv \int \frac{d^{3}\bp}{(2\pi)^3 p^0}$ with $p^0=\sqrt{|\bp|^2+\mass^2}$; $\mass$ denotes the particle mass. Here, $\rP_{(n)}(\dots)$ denotes the $n$-particle reduced distribution function, and $\{\bph\} \equiv (\bx_1,\bp_1, \bx_2,\bp_2, \dots, \bx_n,\bp_n)$ represents the phase-space coordinates of the $n$ particles, with $\bx_i$ and $\bp_i$ indicating the position and momentum of the $i$-th particle, respectively. $\mathcal{W}_{(\bp_i,\bp_j \rightarrow \bp_i',\bp_j')}$ denotes the transition probability density for a pair of particles with initial momenta $\bp_i,\bp_j$ to scatter into final momenta $\bp_i',\bp_j'$. $N$ denotes the total number of particles in the system.

Although the full BBGKY hierarchy is mathematically and computationally formidable, we render it numerically tractable by expanding the distribution functions in an orthonormal, complete basis,
\Eqbasis
where $\Lambda$ has units of energy, $u^\mu$ is the velocity of a reference frame, $Y_{\ell m}$ are the real spherical harmonics, and $L_n^{(2\ell+2)}$ are the Legendre polynomials.

This expansion transforms the BBGKY hierarchy into the sBBGKY hierarchy, which governs the time evolution of the expansion coefficients,
\EqsBBGKY
where the single index $i$ represents a flattened combination of the indices $(n, \ell, m)$ appearing in the basis functions defined in Eq.~\eqref{eq:basis}. Here, $\boldsymbol{B}_{ij}$ denotes the free-streaming integral appearing in the kinetic equation, defined as
\EqinttersotB
where $\basisleft_{n, \ell, m}(p_{\mu})$ are the dual basis functions, defined as
\Eqdualbasis
The basis and the dual-basis functions satisfy the orthogonality condition,
\Eqorthocond
where the integration measure is defined as $\int_{p_{\mu}} \equiv \int d (p_{\mu} u^{\mu} /\Lambda)\, d\Omega$. The three-index tensor $A_{ijk}$ and four-index tensor $C_{ijks}$ on the right-hand side represent collision integrals. The former collects exchanges between a particle in the concern sector and one in the not-in-concern sector (i.e., integrated out), whereas the latter collects exchanges between two concern-sector particles. Both are defined as the difference between the corresponding gain and loss terms,
\EqinttersotATwoterms
\EqinttersotCTwoterms
Explicitly, the gain and loss contributions are given by
\EqinttersotAGain
\EqinttersotALoss
\EqinttersotCGain
\EqinttersotCLoss

\subsection{The Boltzmann Equation}\label{sec:Method:spectral_Boltzmann}

The full (spectral) BBGKY hierarchy provides a systematic, time-reversal-symmetric framework for studying the evolution of many-particle systems in phase space, proceeding step by step from one-body to two-body correlations, and ultimately up to the full $N$-body description. Because each equation in the hierarchy is coupled to the next higher-order equation, solving the entire set remains extremely challenging in practice. Fortunately, lower-order truncations are often sufficient to accurately capture the nonequilibrium dynamics of many-particle systems. In fact, the lowest-order truncation yields the Boltzmann equation, which was introduced even before the BBGKY hierarchy and has become a cornerstone of kinetic theory due to its numerical tractability. Accordingly, we focus on the (spectral) Boltzmann evolution here and defer higher-order correlations to future work.

The resulting leading-order equations of the sBBGKY hierarchy, which form the basis of our analysis, are given by,
\begin{align}\label{eq:P1_SBBGKY}
\begin{split}
\condamp
    \frac{\partial}{\partial t} 
    \rP^{i}(t, \bx)  
+\;
    \boldsymbol{B}_{ij}
    \cdot\frac{\partial }{\partial \bx}
    \rP^{j}(t, \bx)   
\condeq
    (N-1)
    A_{i j k}
    \rP^{j k}(t, \bx, \bx)   \,,
\end{split}
\end{align}
where $\rP^{i}(t, \bx)$ and $\rP^{jk}(t, \bx, \bx)$ respectively represent the spectral expansion coefficients of the single-particle and two-particle probability densities. The quantity $N$ denotes the total number of particles in the many-particle system. We then truncate the many-body distribution and retain only the single-particle distribution. Under the assumption of molecular chaos, this leads to
\begin{align}
    \rP^{i}
=\;& 
    \frac{f^i(t, \bx)}{{N}}\,,
    \label{eq:Define_fg_1}
    \\
\begin{split}
    \rP^{ij}
=\;& 
    \frac{f^i(\bx_1)f^j(\bx_2)}{N^2}
    \,.
\end{split}\label{eq:Define_fg_2}
\end{align}
Substituting Eqs.~\eqref{eq:Define_fg_1}-\eqref{eq:Define_fg_2} into the leading-order equation of the spectral BBGKY hierarchy, Eq.~\eqref{eq:P1_SBBGKY}, yields the spectral Boltzmann equation,
\begin{align}\label{eq:f_SBBGKY}
\begin{split}
\condamp
    \partial_t
    f_{i}(t,\bx)
    +
    \boldsymbol{B}_{ij}
    \cdot
    \partial_{\bx}
    f_j(t, \bx)
\condeq
    \frac{N-1}{N}
    A_{ijk}
    f_{j}(t,\bx)
    f_{k}(t,\bx)
    \;.
\end{split}
\end{align}
This equation serves as the foundation for the analysis presented in this work.

Compared to the conventional Boltzmann equation, the spectral formulation in Eq.~\eqref{eq:f_SBBGKY} accurately captures nonlinear dynamics at a computational cost comparable to conventional linearized methods (see~\cite{Mendoza:2009gm, Romatschke:2011hm}). Specifically, the essential nonlinear physics can be well described by evolving only the first few spectral coefficients defined in (3+1)-dimensional spacetime. The theoretical foundation for this efficiency is discussed in Section~II of Ref.~\cite{Lu:2025yry}, while numerical validation is provided in Section~IV. Moreover, the analytic method for evaluating collision integrals proposed in Ref.~\cite{Lu:2025yry} achieves high numerical accuracy and eliminates the need for ensemble averaging, which is typically required by stochastic methods starting from identical initial conditions. 

In numerical implementations, errors arise only from time integration and spectral truncation. The former can be controlled to fourth-order accuracy using a fourth-order Runge--Kutta scheme, while the latter achieves an accuracy on the order of $\exp(-M)$ when the spectral BBGKY method is applied to smooth distributions~\cite{Press2007}. 

\subsection{Specific Setup and Conventions}\label{sec:Method:notation}

With the computational framework established, we are now ready to study the approach to thermal equilibrium with the full, nonlinear collision kernel. In the present study, we focus on a system subject to the following specific conditions.

We restrict our attention to spatially homogeneous systems, so the spectral coefficients $f_i$ are independent of the coordinate $\boldsymbol{x}$. For distributions with nontrivial spatial dependence, solving the spectral Boltzmann equation amounts to solving a $(3+1)$-dimensional system of hydrodynamic-type equations with $M$ independent fields, where $M$ is the number of basis functions retained by the truncation. Although such simulations are feasible with current hardware, they remain computationally demanding; therefore, we defer a full inhomogeneous study to future work. We expect that our conclusions regarding thermalization will remain valid—at the level of spatial averages—even when homogeneity is relaxed.

We employ a special two-body ($2 \to 2$) scattering kernel to specify interactions,
\begin{align}\label{eq:example_collision_kernel}
W_{(p_1^\mu, p_2^\mu \to p_{3}^{\mu}, p_{4}^{\mu})} = \frac{P^2}{\Lambda^2} \, \sigma^{(0)}(P/\Lambda),
\end{align}
where $P^2 = (p_1^\mu+p_2^\mu)(p_{1,\mu}+p_{2,\mu})$ denotes the invariant center-of-mass energy squared, with the metric convention $g^{\mu\nu} = \mathrm{diag}(+1, -1, -1, -1)$. Moreover, we assume a momentum-independent, isotropic cross section $\sigma^{(0)}(P/\Lambda) \equiv \sigma_0 = \text{const}$. We further restrict our consideration to massless particles. These choices afford simplicity: specifically, the collision integrals $A_{ijk}$ can be evaluated exactly in closed analytic form.  The results are given in Eq.~(95) of Ref.~\cite{Lu:2025yry}.

Finally, for clarity, we summarize the notation adopted herein.

\begin{itemize}
\item As a technical convention, in addition to the system assumptions above, we set the basis scale equal to the equilibrium temperature, $\Lambda = T$, which simplifies equilibrium expressions (see Eq.~(26) of Ref.~\cite{Lu:2025yry}).
\item $t_{\mathrm{lin}}$: the time at which a system evolving under the nonlinear Boltzmann equation enters the linear regime, i.e., when the linearized Boltzmann description becomes applicable.
\item $t_{\mathrm{therm}}$: the time at which the same system reaches thermal equilibrium.
\item $\tau_{\mathrm{lin}}$: Characteristic timescale for linearized evolution, defined by an exponential fit to the difference between the ensembles of distribution functions from nonlinear and linearized evolutions.
\item $\tau_{\mathrm{therm}}$: Characteristic timescale for thermalization, defined by an exponential fit to the ensemble of distribution functions.
\item $t_{\mathrm{match}}$: A matching time parameter used to connect the nonlinear and linearized evolution stages. This parameter is chosen such that $t_{\mathrm{match}} > t_{\mathrm{lin}}$.
\item $(n_{\mathrm{max}}, \ell_{\mathrm{max}})$: Truncation parameters specifying the cutoff in the spectral expansion. For a given $(n_{\mathrm{max}}, \ell_{\mathrm{max}})$, the total number of basis functions is $M = (n_{\mathrm{max}}+1)(\ell_{\mathrm{max}}+1)^2$.
\end{itemize}
\section{Thermalization and Linearization with Full Collision Kernel}\label{sec:Result}

Now we are ready to solve the spectral Boltzmann equation~\eqref{eq:f_SBBGKY} with the full, nonlinear collision kernel, achieving high numerical accuracy and efficiency. We focus in particular on \textit{when a system enters a regime that can be approximated by linearized equations---before it reaches thermal equilibrium}. This is essential for the foundation of hydrodynamics. In what follows, we refer to this process as \textit{linearization}.

We note two special cases: isotropic ($\ell_{\max}=0$) and axisymmetric, angular-only ($n_{\max}=0$ and $m=0$). In both, the third-rank collision tensor $A_{ijk}$ exhibits a recursive structure that permits analytic evolution of the spectral coefficients, thereby providing straightforward intuition and eliminating both time-integration and spectral-truncation errors. Exact solutions for these two simplified scenarios will be discussed in Section~\ref{sec:nonlinear_boltzmann:isotropic} and Section~\ref{sec:nonlinear_boltzmann:anisotropic}. Afterward, we move on to numerical solutions with sufficiently large truncation parameters $n_{\max}$ and $\ell_{\max}$ in Section~\ref{sec:nonlinear_boltzmann:numerical}, thereby treating generic distribution functions with fully coupled radial and angular degrees of freedom.

\subsection{Analytical study of isotropic distributions}\label{sec:nonlinear_boltzmann:isotropic}

We begin with distributions isotropic in momentum space ($\ell_{\max}=0$), i.e., the angular dependence is fixed to its equilibrium form while only radial modes evolve. The spectral coefficients of the Boltzmann equation simplify to $f_n \equiv f_{(n,0,0)}$ for $n \in \{0, \cdots, M-1\}$. The nonlinear equations of motion (EoMs) for the spectral coefficients in the isotropic truncation form a recursively solvable hierarchy, which admits closed-form analytic solutions for each $f_n$. An analytic solution with a specific initial condition has been found in Ref.~\cite{Bazow:2015dha}. Here we extend it to an analytic calculation of the evolution of the coefficients, under proper truncation, with arbitrary initial conditions.

Under the isotropic assumption, the full nonlinear EoMs are given by
\begin{align}
    \partial_t f_n = \frac{1}{c_0\,\tau} \Big( \frac{1}{n+1} \sum_{j=1}^{n-1} f_j \, f_{n-j} - \frac{n-1}{n+1} f_0 \, f_n\Big), \quad \forall\, n\geq 2,
    \label{eq:spectral_boltzmann_isotropic}
\end{align}
where we have defined $c_0 \equiv 2\sqrt{\pi}$ and introduced a relaxation timescale $\tau \equiv \frac{\pi^3}{T^3\,\sigma_0}$. The hierarchy is recursive: the EoM for the $n$th coefficient only involves $f_{j \leq n}$, so we can solve the hierarchy equation by equation and obtain an analytic solution for each $f_n$. The first two spectral modes, $f_{0}=2\sqrt{\pi}$ and $f_{1}=0$, are fixed by conservation of particle number and energy. At the lowest nontrivial order, $f_{2}(t)$ obeys a linear decay, $\partial_t f_2 = -\frac{f_2}{3\tau}$, yielding the standard exponential relaxation to zero; $f_{3}(t)$ satisfies an analogous equation with a different relaxation rate. Starting from $n=4$, the evolution equation becomes a linear relaxation equation with nonhomogeneous terms\footnote{In the context of ordinary differential equations, nonhomogeneous terms refer to terms that do not depend on the unknown function or its derivatives.} contributed by lower-order coefficients. For example, the equation for $f_4$ reads $\partial_t f_4 = -\frac{3}{5}\frac{f_4}{\tau} + \frac{f_2^2}{5 c_0 \tau}$, where the nonhomogeneous term $\frac{f_2^2}{5 \, c_0 \, \tau}$ arises from the lower-order spectral coefficient $f_2$. Analytic solutions for higher-order $f_n$ can be obtained by straightforward---though tedious---solution of the hierarchy equations~\eqref{eq:spectral_boltzmann_isotropic}. For clarity, we list the exact solutions for the first six coefficients:
\begin{align}
\begin{split}
    f_2(t) =\;& c_2\, e^{-\frac{1}{3}\frac{t}{\tau}}\,,
\\
    f_3(t) =\;& c_3\, e^{-\frac{1}{2}\frac{t}{\tau}}\,,
\\
    f_4(t) =\;& c_4\, e^{-\frac{3}{5}\frac{t}{\tau}}
    +\frac{3c_2^2}{c_0} \Big(e^{-\frac{3}{5}\frac{t}{\tau}} - e^{-\frac{2}{3}\frac{t}{\tau}}\Big) \,,
\\
    f_5(t) =\;& c_5\, e^{-\frac{2}{3}\frac{t}{\tau}} + \frac{2c_2c_3}{c_0} \Big(e^{-\frac{2}{3}\frac{t}{\tau}} - e^{-\frac{5}{6}\frac{t}{\tau}}\Big) \,,
\\
    f_6(t) =\;& c_6\, e^{-\frac{5}{7}\frac{t}{\tau}} + \frac{c_0c_3^2-6c_2^3}{2c_0^2}\Big(e^{-\frac{5}{7}\frac{t}{\tau}} - e^{-\frac{t}{\tau}}\Big)
\condbreak{+}
    \frac{30(c_0c_2c_4+3c_2^3)}{23c_0^2}\Big(e^{-\frac{5}{7}\frac{t}{\tau}} - e^{-\frac{14}{15}\frac{t}{\tau}}\Big)\,,
    \label{eq:spectral_boltzmann_isotropic_solutions_nonlinear}
\end{split}
\end{align}
where $c_n \equiv f_n(t=0)$ are the corresponding initial conditions.

To study the linearization of the system, the linear EoMs that correspond to Eq.~\eqref{eq:spectral_boltzmann_isotropic} are needed. They can be obtained by expanding the isotropic spectral Boltzmann equation about thermal equilibrium, i.e., $f_0 = 2\sqrt{\pi}$ and $f_{n \ge 1} = 0$. We may describe the near-equilibrium evolution by retaining only linear terms in $f_{n \ge 1}$. The linearized Boltzmann equation and its exact solution are given by
\begin{align}
\begin{split}
\condamp
    \partial_t f_n^\mathrm{lin} = -\frac{1}{\tau}\frac{n-1}{n+1} f_n^\mathrm{lin}, 
\condbreak{\quad \Rightarrow\quad}
    f_n^\mathrm{lin}(t) = c_n^\mathrm{lin} \, e^{-\frac{n-1}{n+1}\frac{t}{\tau}}\,,
    \quad\forall\, n\geq 2.
    \label{eq:spectral_boltzmann_isotropic_linear}
\end{split}
\end{align}
For nonconserved modes ($n \ge 2$), $f_2$ has the longest relaxation time, $3\tau$; as $n$ increases, the relaxation time, $\frac{n+1}{n-1}\tau$, decreases and approaches $\tau$ for sufficiently large $n$. By comparing the full nonlinear solutions in Eq.~\eqref{eq:spectral_boltzmann_isotropic_solutions_nonlinear} with the linearized ones in Eq.~\eqref{eq:spectral_boltzmann_isotropic_linear}, we find that all coefficients with $n \ge 4$ can be expressed as a superposition of the linear solution and additional corrections originating from nonhomogeneous lower-order terms. The nonhomogeneous contribution to the equation of motion for $f_n$ always takes the form $f_j f_{n-j}$. Such terms decay faster than the linear component, since their decay rate satisfies $\Big(\tfrac{j-1}{j+1} + \tfrac{n-j-1}{n-j+1}\Big)\tfrac{1}{\tau} \;>\; \tfrac{n-1}{n+1}\tfrac{1}{\tau}$. Therefore, all modes undergo a linearization stage before complete thermalization is achieved, indicating that linearized dynamics universally precede equilibration.

A key question is when the linear solution becomes a good approximation---the basic requirement in the construction of hydrodynamic equations as demonstrated earlier. We note that the asymptotic dynamics of $f_n$ [Eq.~\eqref{eq:spectral_boltzmann_isotropic_solutions_nonlinear}] coincide with the linearized evolution~\eqref{eq:spectral_boltzmann_isotropic_linear} but with a shifted effective initial state. For example, the late-time behavior of $f_4$ reads
$f_4(t) = \big(c_4 + \frac{3c_2^2}{c_0}\big)e^{-\tfrac{3}{5}\,\tfrac{t}{\tau}}$,
which reproduces the expected linear exponential decay with an effective initial coefficient $c_4 + 3c_2^2/c_0$ rather than $c_4$. Consequently, the nonlinear evolution should not be compared with the linearized evolution using identical initial conditions. Rather, we determine the coefficients $c_n^{\mathrm{lin}}$ in each linearized evolution~\eqref{eq:spectral_boltzmann_isotropic_linear} so that $f_n^{\mathrm{lin}}$ matches the nonlinear distribution at a sufficiently late time $t_{\mathrm{match}}$, i.e., $f_n^{\mathrm{lin}}(t_{\mathrm{match}})=f_n(t_{\mathrm{match}})$. This provides a quantitative diagnostic for assessing the differences between nonlinear and linearized evolution.

We compute the linearization timescales and compare them with the thermalization ones for two classes of observables: the bulk-pressure–related scalar $\mathcal{M}_{\Pi}$, chosen for its direct connection to hydrodynamic variables, and the energy moments $\mathcal{M}_s$ ($0\le s\le M$), which provide complementary diagnostics of the high-energy tail. For isotropic momentum distributions, angular-dependent observables (e.g., the diffusion current and the shear-stress tensor) vanish, so we focus on the moment that is proportional to the bulk pressure,
\begin{align}
\begin{split}
\mathcal{M}_{\Pi}(t) \equiv\;&
     \frac{4\pi^{\frac{5}{2}}}{T^2}\int \frac{d^3 \bp}{(2\pi)^3 p^0}
     \Big(
     f(t,\boldsymbol{p}) - f_{\mathrm{eq}}(t,\boldsymbol{p})
     \Big)
=
    \sum_{n=1}^{\infty} f_n(t)\,.
\end{split}
\end{align}
The energy moments $\mathcal{M}_s$ are given by
\begin{align}\label{eq:energy_moment_isotropic}
\begin{split}
\mathcal{M}_{s}(t) \equiv\;\condamp
     \frac{4\pi^{5/2}}{(s+2)!  T^{s+3}} 
     \int \frac{d^3 \bp}{(2\pi)^3}
     f(t,\boldsymbol{p}) (p^0)^s
\condeq
    \sum_{n=0}^{s}     
    \frac{(-1)^{n}s!}{(s-n)!n! }
     f_{n}\,.
\end{split}
\end{align}
The energy moments can be evaluated exactly not only because the relevant coefficients admit closed-form solutions, but also because for $s\ge 0$, each $\mathcal{M}_s$ depends only on coefficients $f_m$ with $m \le s$.

We are now ready to study thermalization and linearization---by comparing the nonlinear and linearized Boltzmann evolutions---of $\mathcal{M}_{\Pi}$ and $\mathcal{M}_s$, and we present the results in Figure~\ref{fig:LB_vs_B}. The truncation parameter is set to $n_{\mathrm{max}} = 10$.

\begin{condfigure}
    \centering
    \includegraphics[width=1\textwidth]{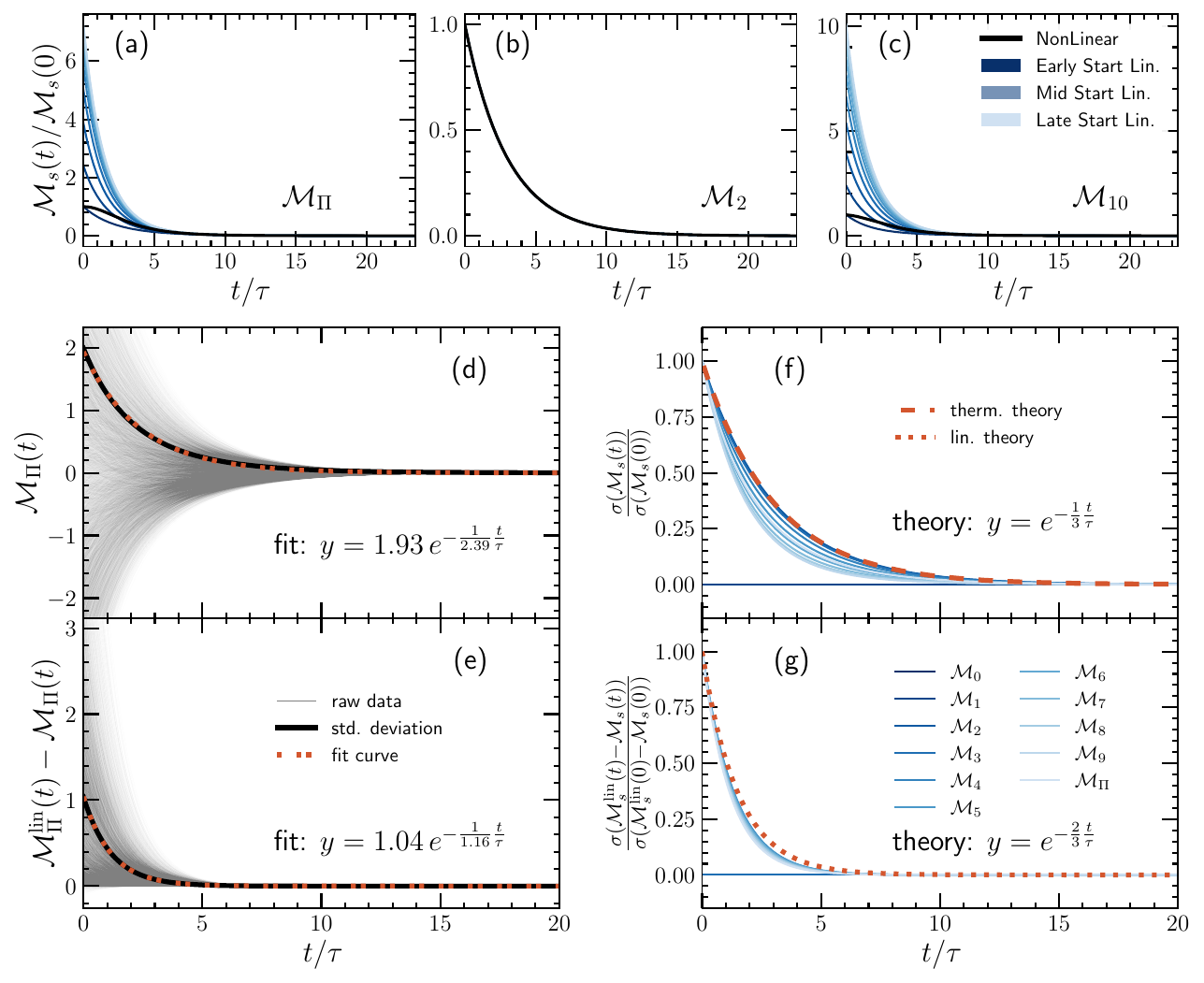}
    \caption{
    (a--c) Full nonlinear solutions (thick black curves) versus linearized approximations (blue curves) for $\mathcal{M}_{\Pi}$, $\mathcal{M}_{2}$, and $\mathcal{M}_{10}$ under a specific initial state. 
    Blue curves from dark to light correspond to linearized solutions matched to the full solution at $t = t_{\mathrm{match}} \in {0, 0.1, \cdots, 0.9}\, t_{\mathrm{max}}$, where $t_{\mathrm{max}} = 23 \, \tau$. 
    Apart from the first few dark-blue trajectories, the subsequent lighter ones merge as the system enters the linearized regime. 
    (d, e) Time evolution of $\mathcal{M}_{\Pi}(t)$ (d) and its deviation from the corresponding linearized solution, $\mathcal{M}_{\Pi}^{\mathrm{lin}}(t) - \mathcal{M}_{\Pi}^{}(t)$ (e). 
    Each gray curve denotes one sampled initial state generated according to Eq.~\eqref{eq:ensemble_1}; in total, $10^{5}$ gray trajectories are shown. 
    At larger $t/\tau$, overlap makes the gray appear black. 
    The standard deviations ($\sigma$) and the corresponding best-fit curves using $y = b\exp[-t/(a \tau)]$ are shown as solid black and dash-dotted orange curves, respectively. 
    (f, g) Normalized ensemble standard deviations, $\sigma(\mathcal{M}_s(t))/\sigma(\mathcal{M}_s(0))$ for $0\le s\le 9$, and normalized deviations from the linearized results, $\sigma(\mathcal{M}^{\mathrm{lin}}_s(t)-\mathcal{M}^{}_s(t))/\sigma(\mathcal{M}^{\mathrm{lin}}_s(0)-\mathcal{M}^{}_s(0))$, are shown as blue solid curves; orange curves indicate the corresponding theoretical upper bounds given in Eq.~\eqref{eq:spectral_boltzmann_isotropic_solutions_nonlinear}.
    }
    \label{fig:LB_vs_B}
\end{condfigure}

To illustrate nonlinear evolution in a many-particle system, we first consider a representative initial state. We initialize with coefficients $\{c_n\}=\{2\sqrt{\pi}, 0, 5, 2.5, 2.5, -1.38\times10^{-2}, -4.32\times10^{-3}, -1.29\times10^{-3}, -3.78\times10^{-4}, -1.08\times10^{-4}, -3.04\times10^{-5}\}$ and track the time evolution of three observables—the bulk-pressure moment $\mathcal{M}_{\Pi}$, a lower-order moment $\mathcal{M}_2$, and a higher-order moment $\mathcal{M}_{10}$. These are displayed in Figure~\ref{fig:LB_vs_B} (a-c), respectively. Each panel shows the nonlinear evolution (black curve) governed by Eq.~\eqref{eq:spectral_boltzmann_isotropic}, alongside a family of linearized evolutions (blue curves) governed by Eq.~\eqref{eq:spectral_boltzmann_isotropic_linear}. We adopt the aforementioned linearization scheme under the assumption that subsequent linear and nonlinear evolutions should agree for $t > t_\mathrm{match}$, and we vary the matching time $t_\mathrm{match} = \{0.0, 0.1, \cdots, 0.9\} \, t_\mathrm{max}$, encoded by the blue gradient in the plots.\footnote{In principle, for the cases in Sections~\ref{sec:nonlinear_boltzmann:isotropic} and~\ref{sec:nonlinear_boltzmann:anisotropic}, the evolution of $f_n$ can be solved analytically, allowing the late-time linearized initial states to be obtained in closed form. For instance, for $f_4$ the corresponding linearized initial condition $c_4^{\mathrm{lin}}$ obtained by matching at $t_{\mathrm{match}}$ is
$$
    c_4^{\mathrm{lin}}(t_{\mathrm{match}}) = c_4
    +\frac{3c_2^2}{c_0} \Big(1- e^{-\frac{1}{15}\frac{t_{\mathrm{match}}}{\tau}}\Big) \,.
$$
From this expression, the limiting value of $c_4^{\mathrm{lin}}$ as $t_{\mathrm{match}} \to \infty$ can be derived straightforwardly. In practice, however, we evaluate $c_4^{\mathrm{lin}}$ at a finite $t_{\mathrm{match}}$ to maintain consistency with Section~\ref{sec:nonlinear_boltzmann:numerical}, where the linearized initial condition cannot be obtained analytically and a finite matching time is unavoidable.} Here, $t_\mathrm{max} = 23\tau$ is defined as the time at which all nonconserved modes---$f_n$ with $n \geq 2$---become sufficiently small, indicating thermal equilibrium.

As expected from the fact that $\mathcal{M}_2(t)$ evolves independently and obeys a closed linear equation~\eqref{eq:spectral_boltzmann_isotropic_solutions_nonlinear}, it remains identical to its linear counterpart throughout the evolution. In contrast, both $\mathcal{M}_{\Pi}$ and $\mathcal{M}_{10}$ exhibit clear nonlinear behavior at early times and gradually align with the linearized evolution. When $t_\mathrm{match}$ is small (e.g., $0.0\, t_\mathrm{max}, 0.1\, t_\mathrm{max}$), the linearized curves deviate visibly from the nonlinear evolution for $t>t_{\mathrm{match}}$, and the corresponding initial conditions differ markedly. For $t_\mathrm{match} \geq 0.2\, t_\mathrm{max}$, the linearized and nonlinear curves nearly overlap for $t > t_\mathrm{match}$, indicating that the system has entered the linear regime by $t \approx 0.2\, t_\mathrm{max}$. Moreover, the initial conditions of these late-time linearized evolutions converge to a common form, which implies that once the system enters the linear regime, the corresponding linearized initial condition can be consistently reconstructed for any choice of $t_\mathrm{match}$ greater than that time.

Results in Figure~\ref{fig:LB_vs_B} (a--c) demonstrate that, under isotropic conditions, a nonlinear system inevitably transitions into a linear regime before it approaches thermal equilibrium. The evolution in the linear regime can be mapped to that of a system that starts from a different initial condition and evolves according to the linearized Boltzmann equation from the outset---regardless of whether this initial condition has actually entered the linear regime---denoted as ``IC lin.'' By quantifying the deviation between the nonlinear evolution curve and the ``IC lin'' curve, we can determine when the system enters the linear regime. While this phenomenon is observed for a specific initial distribution, we will show below that it is a universal feature that holds in general.

While the foregoing example is instructive, the significant event-by-event variability of initial distributions in heavy-ion collisions necessitates a statistical description of the initial state itself---namely, a distribution-function ensemble. The Boltzmann equation inherently encodes the ensemble-averaged behavior as typically understood in statistical mechanics. However, the uncertainty in the initial state requires a different kind of ensemble average: one over the possible initial one-particle distribution functions. We refer to this concept---which goes beyond the traditional ensemble averaging in statistical mechanics---as a ``distribution function ensemble.''

To assess the generality of nonlinear evolution, we construct an ensemble of $10^5$ random initial distribution functions. Each set of expansion coefficients $\{ c_n \}$ is sampled from a uniform distribution with compact support
\begin{align}
|c_n| < 2\, e^{-(n/3) \, \Theta(n - 4.5)},\quad n\ge2,
\label{eq:ensemble_1}
\end{align}
where $\Theta(n)$ is the Heaviside step function, defined as $\Theta(n) = 1$ for $n \geq 0$ and $\Theta(n) = 0$ otherwise. This choice ensures that low-order modes ($n < 5$) remain unrestricted, while higher modes are exponentially suppressed, reflecting typical smooth initial distributions and thereby justifying a truncation at $n_{\max}=10$. We set $c_0 = 2\sqrt{\pi}$ and $c_1 = 0$ to fix the total particle number and energy. Each sample is evolved using the full nonlinear equation and compared with its corresponding linearized trajectory, matched at $t_\mathrm{match} = 10\,\tau$.

We first examine the ensemble evolution of $\mathcal{M}_\Pi(t)$ [see Figure~\ref{fig:LB_vs_B}(d)]. Each gray curve corresponds to a single sample, while the thick black curve represents the standard deviation across the ensemble.\footnote{We focus on the standard deviation, rather than the mean, because the initial conditions are uniformly sampled around zero, and the mean of the distribution-function ensemble offers no meaningful insight.} The decay of the standard deviation is well described by an exponential fit, $y \propto e^{-t / (2.39 \tau)}$, shown as the dashdot orange curve. For comparison, the slowest theoretical decay mode scales as $e^{-t / (3 \tau)}$, which corresponds to the slowest term in the analytic solution, Eq.~\eqref{eq:spectral_boltzmann_isotropic_solutions_nonlinear}. This indicates that $\mathcal{M}_\Pi(t)$ thermalizes slightly faster than the theoretical lower bound on average.

We then quantify the ensemble-level deviation between the nonlinear evolution and its linearized counterpart, $\mathcal{M}_\Pi^{\mathrm{lin}}(t)\equiv \sum_{n=1}^{\infty} f_n^{\mathrm{lin}}(t)$ [see Figure~\ref{fig:LB_vs_B}(e)]. As before, individual samples are shown as gray lines, and the black curve tracks the standard deviation across the ensemble. This deviation decreases exponentially, with a best-fit curve $y \propto e^{-t / (1.16 \tau)}$ shown as a dashdot orange line. In a statistical sense, linearization occurs on a timescale of $\tau_{\mathrm{lin}} = 1.16\tau$, which is shorter than the thermalization time $\tau_{\mathrm{therm}} = 2.39\tau$. It is also slightly shorter than the theoretical upper bound, $3\tau/2$, which corresponds to the slowest nonhomogeneous term in the analytic solution, Eq.~\eqref{eq:spectral_boltzmann_isotropic_solutions_nonlinear}.

For the remaining moments $\mathcal{M}_s$, thermalization and linearization are assessed via the ensemble standard deviations of the moments and of the nonlinear-linear differences, with results summarized in Figure~\ref{fig:LB_vs_B}(f, g). For thermalization, we observe that $\sigma(\mathcal{M}_0) = \sigma(\mathcal{M}_1) = 0$ due to conservation laws. All other observables exhibit exponential decay and approach thermal equilibrium at similar times, with all thermalization times no larger than the theoretical prediction for the slowest mode (dashed orange curve). For the differences between $\mathcal{M}_s$ and their linear counterparts [$\mathcal{M}_{s}^\mathrm{lin}(t) \equiv \sum_{n=0}^{s} \frac{(-1)^{n} \, s!}{n! (s-n)!} f_n^\mathrm{lin}(t)$], $s=2$ and $3$ exhibit no deviations, as these moments depend only on modes governed by linear dynamics. For all other moments, the standard deviation across the distribution function ensemble decays exponentially, approaching zero with a characteristic timescale bounded above by $3\tau/2$ (orange dotted line). We also observe a modest, monotonic, negative $s$-dependence in both $\tau_{\mathrm{therm}}^{[\sigma(\mathcal{M}_s)]}$ and $\tau_{\mathrm{lin}}^{[\sigma(\mathcal{M}_s)]}$, which is opposite to those obtained using formal solutions of the Boltzmann equation that assume a \textit{linearized} collision kernel based on relaxation time approximation~\cite{Strickland:2018ayk}.
   
\begin{condfigure}
    \centering
    \includegraphics[width=1\textwidth]{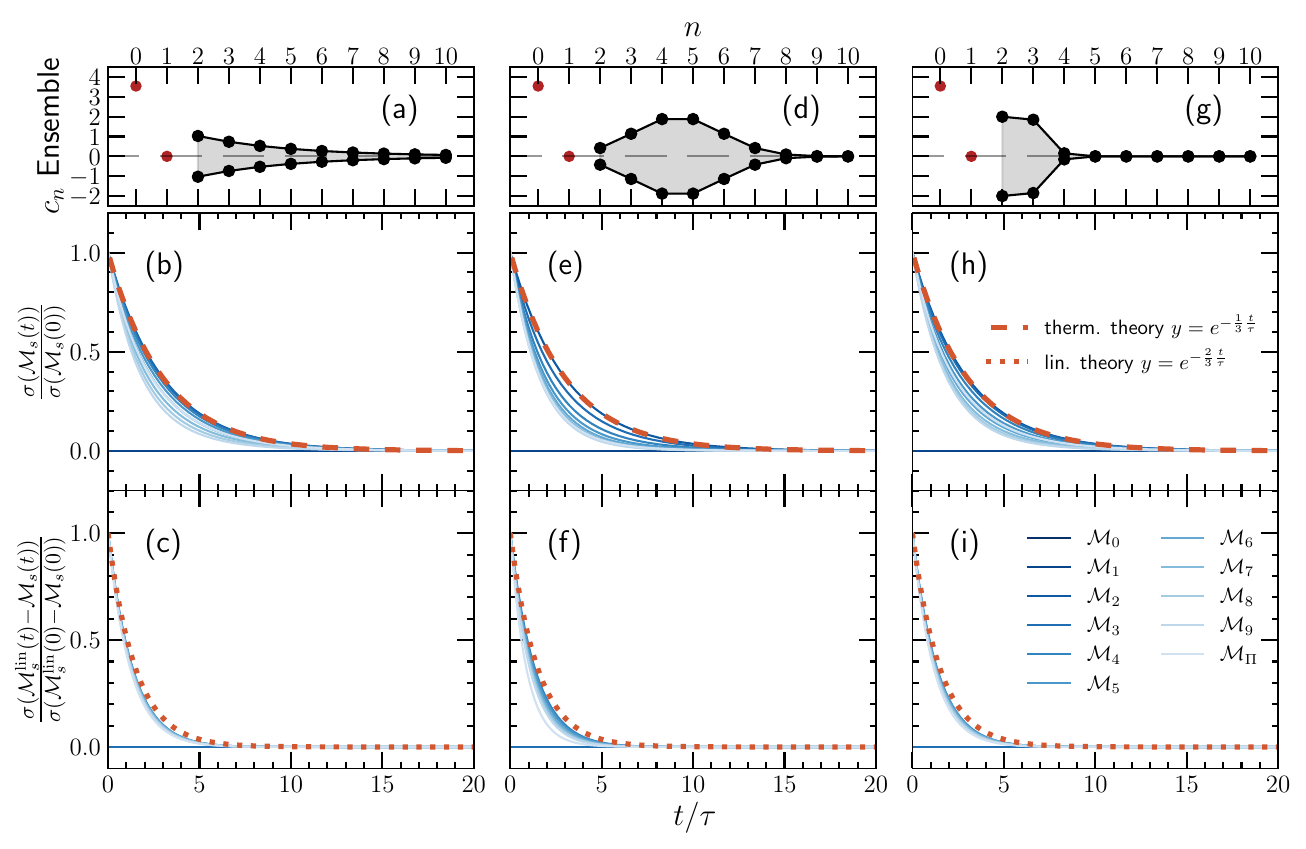}
    \caption{
    Similar to Figure~\ref{fig:LB_vs_B} (f, g), but for three distinct ensembles. 
    Panels (a), (d), and (g) display the admissible ranges of spectral coefficients $c_n$ used to generate the ensembles defined by Eqs.~\eqref{eq:ensemble_2} (exponential), \eqref{eq:ensemble_3} (Gaussian), and \eqref{eq:ensemble_4} (smoothed Heaviside), respectively. Coefficients $c_0$ and $c_1$ are fixed by conservation/Landau matching (red markers), while the allowed ranges for $c_n$ with $n\ge2$ are indicated by gray bands; black markers denote the bounds.
    Thermalization and linearization results are shown in panels (b, c) for exponential, (e, f) for Gaussian, and (h, i) for smoothed Heaviside ensembles.
    \label{fig:Sfig_LB_vs_B_isotropic}}
\end{condfigure}

Both the thermalization time and the linearization time are found to be robust with respect to the choice of the distribution ensemble~\eqref{eq:ensemble_1}. We demonstrate this by performing the same analysis across three additional ensembles of initial distribution functions: an exponential distribution, 
\begin{align}\label{eq:ensemble_2}
|c_n| < 2 \, e^{-n/3}, \quad n \ge 2;
\end{align}
a Gaussian distribution,
\begin{align}\label{eq:ensemble_3}
|c_n| < e^{-(n - 4.5)^2 / 4}, \quad n \ge 2;
\end{align}
and a smoothed Heaviside distribution,
\begin{align}\label{eq:ensemble_4}
|c_n| < \frac{2}{e^{5(n - 4.5)} + 1}, \quad n \ge 2.
\end{align}
Each ensemble consists of $10^5$ samples of initial distribution functions. The corresponding results are shown in Figure~\ref{fig:Sfig_LB_vs_B_isotropic}, panels (a–c), (d–f), and (g–i), respectively. In all three ensembles, thermalization proceeds at a rate no slower than the theoretical bound $e^{-t/(3\tau)}$, and linearization occurs no slower than $e^{-2t/(3\tau)}$. Across all four ensembles, the thermalization timescales of all observables are broadly similar, and their linearization times are nearly identical. This implies that both the thermalization and linearization times are insensitive to the statistical properties of the initial distribution ensemble.

In summary, this subsection examines the linearization-thermalization picture using two classes of observables: (i) the bulk-pressure-related scalar $\mathcal{M}_{\Pi}$, selected for its direct connection to hydrodynamic fields, and (ii) the energy moments $\mathcal{M}_s$ ($0\le s\le M$), which provide complementary sensitivity to the high-energy tail. A further advantage of $\mathcal{M}_s$ in our spectral Boltzmann formulation is that the low-order moments admit closed-form evaluation; consequently, the inferred thermalization and linearization dynamics for these observables are exact---independent of numerical truncation and time integration. Across the observables examined, thermalization times are mutually comparable, linearization times are likewise comparable, and a robust ordering emerges: $\tau_{\mathrm{lin}} \approx \tau_{\mathrm{therm}}/2$.

\subsection{Analytical study of axisymmetric (angular-only) distributions}
\label{sec:nonlinear_boltzmann:anisotropic}

In addition to the isotropic case treated in the previous subsubsection, we now consider an axisymmetric angular sector, which also admits an analytic treatment and thus serves as a second benchmark for linearization and thermalization dynamics. The radial dependence is fixed to its equilibrium form, while only angular modes evolve; i.e., we truncate the radial expansion at $n_{\max}=0$ and impose axisymmetry by setting $m=0$.

Under the axisymmetric, angular-only assumption, the full nonlinear EoMs for the spectral coefficients $f_\ell \equiv f_{(0,\ell,0)}$ are given by:
\begin{align}\label{eq:spectral_boltzmann_anisotropic}
\begin{split}
    \Big(c_0\tau\,\partial_t + \frac{5c_0}{9}\Big) f_2 =\;& -\frac{4\sqrt{5}}{9} \, f_2^2\,,\\
    \Big(c_0\tau\,\partial_t + \frac{11c_0}{16}\Big) f_3 =\;& -\frac{3\sqrt{5}}{5}f_2 \, f_3\,,\\
    \Big(c_0\tau\,\partial_t + \frac{19c_0}{25}\Big) f_4 =\;& \frac{3}{175} \, f_2^2 \, -\frac{9}{25} \, f_3^2-\frac{16\sqrt{5}}{35}f_2 \, f_4\,,\\
\end{split}
\end{align}
where $\tau = \frac{\pi^3}{T^3\,\sigma_0}$ as defined in Section~\ref{sec:nonlinear_boltzmann:isotropic}. This formulation produces a recursive hierarchy that can, in principle, be solved sequentially. In practice, however, the analytic solution for $f_4$ is already algebraically complex. Therefore, we restrict our analysis to the solutions for $f_2$ and $f_3$, 
\begin{align}\label{eq:spectral_boltzmann_anisotropic_solution}
\begin{split}
\condamp
    f_2(t)
=
    \frac{c_2}{e^{\frac{5}{9}\frac{t}{\tau}}+\frac{4 \sqrt{5} c_2}{5c_0}  \big(e^{\frac{5}{9}\frac{t}{\tau}}-1\big)}\,,
\condbreak{\qquad}
    f_3(t)
=
  \frac{c_3\, e^{\frac{1}{16}\frac{t}{\tau}}}{\Big( e^{\frac{5 }{9}\frac{t}{\tau}}+\frac{4 \sqrt{5} c_2}{5c_0} \big(e^{\frac{5}{9}\frac{t}{\tau}}-1\big)\Big)^{\frac{27}{20}}}\,,
\end{split}
\end{align}
and use these to study the corresponding thermalization and linearization dynamics. Under this setup($n \le 0, m = 0$), the linear EoMs and their solutions are
\begin{align}\label{eq:spectral_boltzmann_anisotropic_linear}
\begin{split}
\condamp
    \,\partial_t f_n^{\mathrm{lin}} =- \frac{1}{\tau}\frac{n^2+n-1}{(n+1)^2} f_n^{\mathrm{lin}} \,,
\condbreak{\quad \Rightarrow\quad}
    f_n^{\mathrm{lin}} = c_n^{\mathrm{lin}} e^{-\frac{n^2+n-1}{(n+1)^2}\frac{t}{\tau}}\,,
    \quad\forall\, n\geq 2.
\end{split}
\end{align}

We now turn to our main objective: quantifying the linearization and thermalization times.
We follow the statistical procedure outlined in Section~\ref{sec:nonlinear_boltzmann:isotropic}. 

An ensemble of $10^5$ initial distributions is generated subject to the constraints
\begin{align}\label{eq:anisotropic_ensemble}
|c_2| < \frac{c_0 \sqrt{5}}{4}, \quad |c_3| < \frac{c_0 \sqrt{5}}{8}.
\end{align}
The bound on $c_2$ ensures stability of the analytic solution~\eqref{eq:spectral_boltzmann_anisotropic_solution}, while the stricter bound on $c_3$ reflects the expected decay of anisotropy with increasing angular mode number $\ell$. The coefficients $c_0$ and $c_1$ are fixed to $2\sqrt{\pi}$ and $0$, respectively, by the Landau matching conditions.

We then evolve each initial state of the ensemble using the full nonlinear Boltzmann equation. For every nonlinear trajectory, we construct a corresponding linear evolution that best matches its late-time behavior. This procedure follows the same definition of ``matching'' as used in Section~\ref{sec:nonlinear_boltzmann:isotropic}. In this subsection, $t_{\mathrm{match}} = 10 \tau$.

To characterize the system's response, we define anisotropic moments
\begin{align}\label{eq:anisotropic_moment}
L_i(t) = \int \frac{d^3 \bp}{(2\pi)^3} \, (p^z)^i f(t,\bp)\,.
\end{align}
The moment $L_i(t)$ exhibits a bounded dependence on the coefficients $f_{(n, \ell, m)}$ for $n \leq 0$ and $\ell \leq i$. This restriction follows directly from the orthogonality of the spherical harmonics and the associated Laguerre polynomials. Here we focus on $L_2$ and $L_3$ since, under the truncation $n_{\mathrm{max}} = 0, \, \ell_{\mathrm{max}} = 3$ and the constraint $m = 0$, all other moments either vanish or evolve identically to $L_2$ or $L_3$.

The statistical measures of thermalization and linearization for axisymmetric, angular-only modes in Figure~\ref{fig:N0L6_therm_and_lin} exhibit qualitatively similar behavior to the isotropic case in Section~\ref{sec:nonlinear_boltzmann:isotropic}, suggesting a robustness in the approach to equilibrium. In Figure~\ref{fig:N0L6_therm_and_lin}(a), we plot the ensemble standard deviations of $L_2(t)$ and $L_3(t)$; and in Figure~\ref{fig:N0L6_therm_and_lin}(b), we plot the ensemble standard deviation of the difference between the nonlinear evolution and the matched linear solution. In both panels, the data appear as blue curves. Each curve is fitted to $y=\exp[-t/(a \tau)]$, with dashed orange lines indicating the fits (darker for $L_2$, lighter for $L_3$), and the corresponding parameters annotated in the figure. Two systematic trends follow. First, the linearization time ($\tau_{\mathrm{lin}}$) is approximately one half of the thermalization time ($\tau_{\mathrm{therm}}$), consistent with the isotropic case. Second, the moments $L_2$ and $L_3$ exhibit similar thermalization and linearization times, further confirming the universality of the observed behavior across different moments.
\begin{condfigure}
    \centering
    \includegraphics[width=1\textwidth]{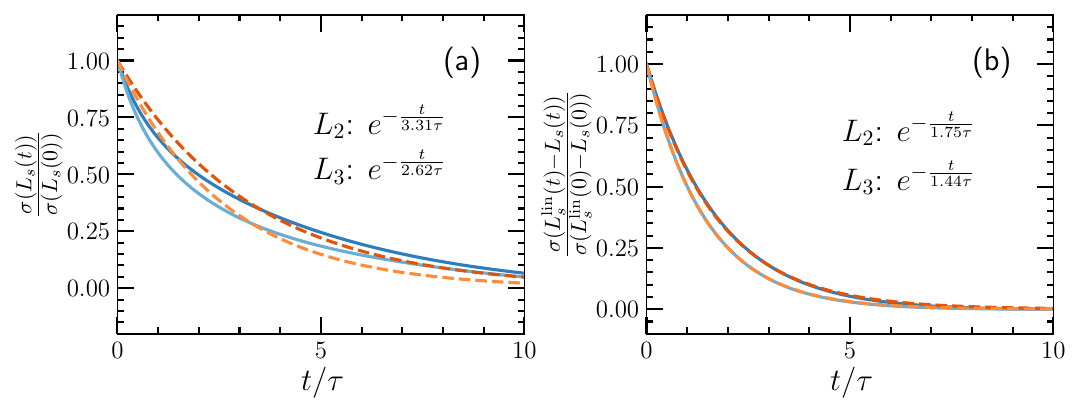}
    \caption{
    Ensemble analysis of spectral-mode evolution in the $n_{\max}=0$ truncation (radial dependence fixed; only angular modes evolve). 
    (a) Normalized ensemble standard deviations $\sigma(L_s(t))/\sigma(L_s(0))$ for $s=2,3$ (blue solid), computed over $10^{5}$ initial distribution functions generated according to Eq.~\eqref{eq:anisotropic_ensemble}. Moments $L_s$ are defined in Eq.~\eqref{eq:anisotropic_moment}. Orange dashed curves are best-fit exponentials, $y=\exp[-t/(a\tau)]$; fit parameters are annotated in the panel.
    (b) For the same ensemble, normalized deviations from the linearized solutions, $\sigma(L^{\mathrm{lin}}_s(t)-L_s(t))/\sigma(L^{\mathrm{lin}}_s(0)-L_s(0))$ for $s=2,3$ (blue solid). Orange dashed curves show best-fit exponentials; fit parameters are annotated in the panel. 
} 
    \label{fig:N0L6_therm_and_lin}
\end{condfigure}
The exponential fit to the thermalization behavior is somewhat less precise in Figure~\ref{fig:N0L6_therm_and_lin}~(a). This modest deviation arises primarily because only a limited number of modes---specifically, $f_2$ and $f_3$---are available analytically. With only these two modes accessible in closed form, even large-scale ensemble averaging cannot fully reproduce the expected exponential decay profile. Nevertheless, the key dynamical features remain clearly visible, and the observed trends are physically consistent. By contrast, the numerical results presented in Section~\ref{sec:nonlinear_boltzmann:numerical} (under the truncation $n_{\mathrm{max}} = 4,\, \ell_{\mathrm{max}} = 4$) and the analytic results in Section~\ref{sec:nonlinear_boltzmann:isotropic} (for $n_{\mathrm{max}} = 10,\, \ell_{\mathrm{max}} = 0$) exhibit excellent agreement with exponential behavior for both linearization and thermalization. These complementary results underscore the validity of the analytic framework while highlighting the benefits of broader mode access in confirming quantitative trends.

In summary, in this subsection we diagnose linearization and thermalization using the anisotropic moments $L_i(t)$. This choice is motivated by two considerations. First, in the $n_{\max}=0$ truncation the moments $L_i$ depend only on spectral modes with $n=0$ and $\ell\le i$, so their evolution can be evaluated exactly (i.e., without numerical truncation or time-integration errors). Second, under the truncation $n_{\max}=0$, $\ell_{\max}=3$, and $m=0$, all other moments either vanish or evolve identically to $L_2$ or $L_3$, making $L_2$ and $L_3$ representative. Analytically, with the radial dependence fixed to its equilibrium form and only angular modes evolving, a robust ordering still emerges: the linearization time is approximately half the thermalization time, $\tau_{\mathrm{lin}}/\tau_{\mathrm{therm}}\approx 0.5$.

\subsection{Numerical simulations for generic distribution function}\label{sec:nonlinear_boltzmann:numerical}

Finally, we assess the robustness of the linearization-thermalization picture for generic initial conditions by retaining both isotropic and anisotropic low-order spectral coefficients in this subsection. We perform this analysis by numerically solving the full, nonlinear spectral Boltzmann equation~\eqref{eq:f_SBBGKY} under the truncation $n_{\max}=4,\ell_{\max}=4$.

The nonlinear EoMs~\eqref{eq:f_SBBGKY} are the most general formulation of the Boltzmann dynamics employed in this paper. The corresponding linearized Boltzmann equation is derived from the full, nonlinear spectral Boltzmann equation by expanding the distribution function around thermal equilibrium and retaining only terms linear in the nonthermal corrections. This approximation leads to the following evolution equation:
\begin{align}
\begin{split}
    \frac{\partial}{\partial t} 
    f_i(t)
+\;
    \boldsymbol{B}_{ij}
    \cdot\frac{\partial }{\partial \bx}
    f_j(t) 
=\;& 
    (N-1)
    A_{i j}^\mathrm{lin}
    f_j(t),
\end{split}
\end{align}
where $A_{ij}^\mathrm{lin} = (A_{ij0} + A_{i0j})f_{(0,0,0)}$ denotes the linearized collision matrix. This linearization procedure automatically enforces conservation of particle number and energy-momentum, since the corresponding components of the linearized collision matrix vanish. 

To study the linearization and thermalization timescales, we adopt an ensemble-based analysis similar to that used in Sections~\ref{sec:nonlinear_boltzmann:isotropic} and ~\ref{sec:nonlinear_boltzmann:anisotropic}. Specifically, we sample $10^3$ random initial distribution functions, evolve each one using the full, nonlinear Boltzmann equation, and construct the corresponding linearized evolution that best matches the nonlinear trajectory at late times. From the evolving distribution functions, we compute moments and extract their standard deviations across the ensemble. The thermalization time $\tau_{\mathrm{therm}}$ is characterized by the decay of each moment's ensemble standard deviation. Similarly, the linearization time $\tau_{\mathrm{lin}}$ is characterized by the decay of the standard deviation of the difference between the nonlinear and linearized moments over time. In this subsection, the approach used to determine the corresponding linearized initial condition (``IC lin'') is conceptually similar to that described in Section~\ref{sec:nonlinear_boltzmann:isotropic} or Section~\ref{sec:nonlinear_boltzmann:anisotropic}, though the technical implementation details differ slightly. In Section~\ref{sec:nonlinear_boltzmann:isotropic} or Section~\ref{sec:nonlinear_boltzmann:anisotropic}, $t_{\mathrm{match}}$ is chosen as a fixed reference time, specifically $t_{\mathrm{match}} = 10 \, \tau$. Since the full time evolution of the spectral coefficients in that case is analytically known, we can ensure that the system is already sufficiently close to the linear regime at that time. Contrastingly, in this subsection, thermalization is determined adaptively within the simulation. The system is considered to have reached equilibrium once all nonconserved components of $f$ fall below a predefined threshold, $f_{\mathrm{threshold}}$. At that point, the simulation is automatically terminated, and the corresponding time is recorded as $t_{\mathrm{therm}}$. We then define $t_{\mathrm{match}} = 0.3 \, t_{\mathrm{therm}}$, motivated by our observation that, for almost all initial conditions, the linear and nonlinear evolutions agree for $t > 0.3 \, t_{\mathrm{therm}}$; in other words, the system has entered the linear regime by then. This approach allows for a consistent identification of $t_{\mathrm{match}}$ even in the absence of analytical control over the full time evolution. We then evolve the spectral coefficients at $t_{\mathrm{match}}$ backward, so that the linearized initial state corresponding to a given nonlinear trajectory can be reconstructed. This permits a direct comparison between the actual nonlinear dynamics and their linearized counterparts, elucidating how the system approaches the linear regime. Throughout this work, we set $f_{\mathrm{threshold}} = 2\times 10^{-3}$.

The ensemble of initial distribution functions is shown in Figure~\ref{fig:NxLx_Moments}~(a). 
Red points denote coefficients fixed by energy-momentum conservation and the Landau matching conditions,
\begin{align}\label{eq:N4L4DistributionEnsemble_main_1}
\begin{split}
\condamp
    f_{(0,0,0)} = 2\sqrt{\pi}, 
\condbreak{\quad} 
    f_{(1,0,0)} = f_{(0,1,0)} = f_{(0,1,1)} = f_{(0,1,-1)} = 0.
\end{split}
\end{align}
Gray points correspond to sampled coefficients uniformly distributed subject to the constraint:
\begin{align}\label{eq:N4L4DistributionEnsemble_main_2}
|f_{(n, \ell, m)}| \le e^{-\ell}.
\end{align}

\begin{condfigure}
    \centering
    \includegraphics[width=1\textwidth]{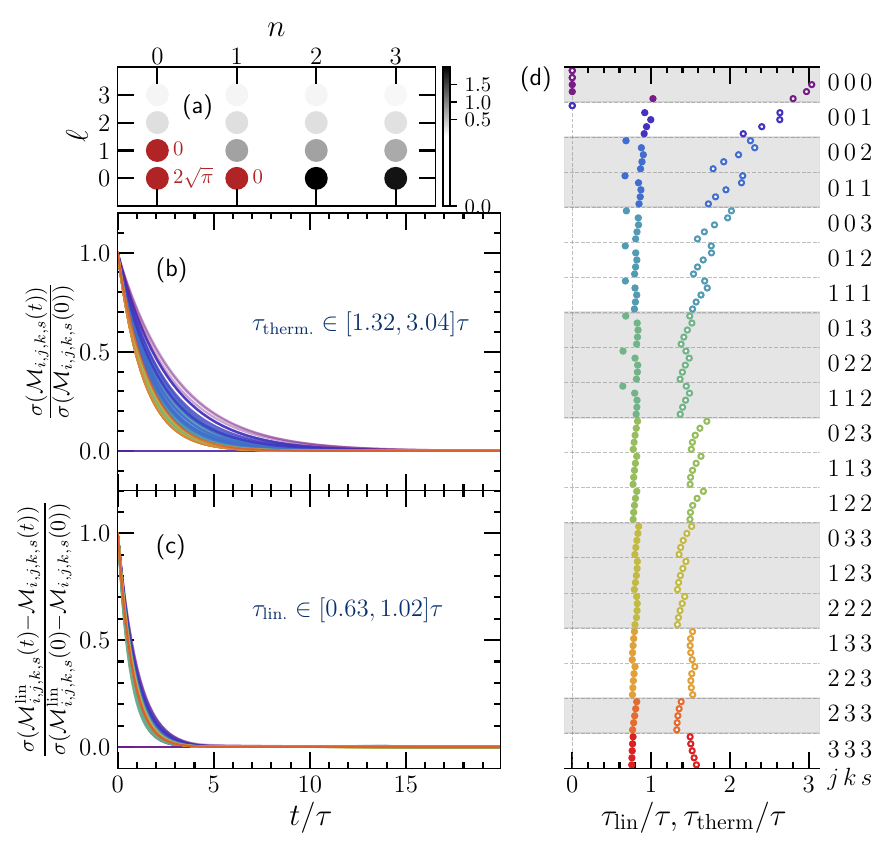}
    \caption{
    Linearization and thermalization times for the moments $\mathcal{M}_{i, j, k, s}$ defined in Eq.~\eqref{eq:NxLx_Moments}, using an ensemble of initial conditions with truncation $(n_{\max},\ell_{\max})=(4,4)$.
    (a) Ensemble specification generated by Eqs.~\eqref{eq:N4L4DistributionEnsemble_main_1}–\eqref{eq:N4L4DistributionEnsemble_main_2}. Red markers indicate coefficients fixed by conservation/Landau matching (values given in Eq.~\eqref{eq:N4L4DistributionEnsemble_main_1}); gray markers denote the absolute values of the endpoints of the uniform sampling ranges for the remaining coefficients, as specified in Eq.~\eqref{eq:N4L4DistributionEnsemble_main_2}. 
    (b) Ensemble-averaged thermalization $\sigma(\mathcal{M}_{i, j, k, s}(t))/\sigma(\mathcal{M}_{i, j, k, s}(0))$ and (c) linearization $\sigma(\mathcal{M}^{\mathrm{lin}}_{i, j, k, s}(t)-\mathcal{M}_{i, j, k, s}(t))/\sigma(\mathcal{M}^{\mathrm{lin}}_{i, j, k, s}(0)-\mathcal{M}_{i, j, k, s}(0))$ of various moments (rainbow lines), each fitted with an exponential decay $\exp[-t/(a\tau)]$. The fastest and slowest decays are indicated in the text.
    (d) Summary of extracted timescales: open markers denote thermalization; filled markers represent linearization. Each horizontally divided region (separated by gray dashed lines) corresponds to a specific $(j, k, s)$ combination (labels at right). Within each band, markers from top to bottom correspond to $i = 0$ to $4$. Colors indicate the value of $j+k+s$.
    }
    \label{fig:NxLx_Moments}
\end{condfigure}

Similar to the previous two subsections, we first examine the $\mathcal{M}_{i, j, k, s}$ moments, defined as
\begin{align}\label{eq:NxLx_Moments}
\mathcal{M}_{i, j, k, s}(t) = \int \frac{d^3\bp}{(2\pi)^3} \, f(t,\bp) \, (p^0)^i (p^x)^j (p^y)^k (p^z)^s,
\end{align}
where $p^\mu$ are the components of the four-momentum. They are the multi-index generalizations of the moments $M_s$ and $L_s$ analyzed earlier, and their dependence on the spectral coefficients $f_{(n, \ell, m)}$ is bounded.

The ensemble behavior of these moments is summarized in Figure~\ref{fig:NxLx_Moments}: panel (b) shows their ensemble-averaged decay, while panel (c) displays the ensemble-averaged deviations between the nonlinear trajectories and their linearized counterparts. To investigate how the thermalization time $\tau_{\mathrm{therm}}$ and the linearization time $\tau_{\mathrm{lin}}$ depend on the indices $(i, j, k, s)$ of each moment, we perform standard exponential fits, $y = \exp[-t/(a \tau)]$, to the ensemble-averaged, scaled decay curves. For reference, the fastest and slowest decay rates across all observables are displayed as text in panels (b) and (c), indicating the minimum and maximum characteristic timescales for thermalization [panel (b)] and linearization [panel (c)] across all moments. All extracted timescales are shown in Figure~\ref{fig:NxLx_Moments}~(d), where filled (open) markers represent $\tau_{\mathrm{lin}}$ ($\tau_{\mathrm{therm}}$). In Figure~\ref{fig:NxLx_Moments}~(d), moments sharing the same energy index $i$ and all permutations of the spatial indices form identifiable groups. Within each group, the average decay time is plotted as a point, and the standard deviation is indicated by an error bar. These error bars are typically too small to be visible, reflecting a high degree of consistency across permutations. For clarity, the grouped spatial indices $(j, k, s)$ are sorted in ascending order of $j+k+s$ and listed in the rightmost column. The energy index $i$ is encoded vertically within the same group: each dashed horizontal band corresponds to a fixed $(j, k, s)$, with increasing $i = 0, 1, 2, 3, 4$ from top to bottom.

The statistical diagnostics of thermalization and linearization in Figure~\ref{fig:NxLx_Moments}~(b–d) closely mirror those reported in Section~\ref{sec:nonlinear_boltzmann:isotropic} and Section~\ref{sec:nonlinear_boltzmann:anisotropic}, indicating a robust pattern in the system's approach to equilibrium. First, the thermalization rates across different moments are broadly similar, and the same holds for the linearization rates. Second, linearization consistently occurs earlier than thermalization for all moments considered (with $\tau_\mathrm{lin}/\tau_\mathrm{therm} \lesssim 1/2$). Third, moments with higher-order energy weights tend to relax more rapidly. Notably, the variation among linearization times is smaller than that among thermalization times, suggesting that the onset of linear behavior is less sensitive to the specific choice of moment than the full thermalization process.

Having established the behavior of the four-momentum moments $\mathcal{M}_{i, j, k, s}(t)$, we now turn to the central hydrodynamic observable—the energy–momentum tensor $T^{\mu\nu}$. The evolution of $T^{\mu\nu}$ determines the hydrodynamic fields (energy density and flow) and the dissipative sector—most notably the shear-stress tensor and bulk pressure—whose relaxation is governed by the shear and bulk viscosities. The energy–momentum tensor $T^{\mu\nu}$ is given by
\begin{align}\label{eq:NxLx_Tmunu}
T^{\mu\nu} = \int \frac{d^3 \bp}{(2\pi)^3 p^0} p^\mu p^\nu f(t,\bp)\,,
\end{align}
with components explicitly mapped to the spectral coefficients as follows,
\begin{align}
T^{tt} =\;& \frac{12 \sqrt{\pi}}{(2\pi)^3} \Lambda^4 \big(f^{(0,0,0)} - f^{(1,0,0)}\big)\,,
\\
T^{tx} =\;& \frac{16 \sqrt{3\pi}}{(2\pi)^3} \Lambda^4 f^{(0,1,1)}\,,
\\
T^{ty} =\;& \frac{16 \sqrt{3\pi}}{(2\pi)^3} \Lambda^4 f^{(0,1,-1)}\,,
\\
T^{tz} =\;& \frac{16 \sqrt{3\pi}}{(2\pi)^3} \Lambda^4 f^{(0,1,0)}\,,
\\
T^{xx} =\;& 
\frac{4\sqrt{\pi}}{(2\pi)^3} \Lambda^4 \Big[
f^{(0,0,0)}-f^{(1,0,0)}
\condbreak{-}
4\sqrt{5} \sum_{n=0}^{\infty} f^{(n,2,0)}
+4\sqrt{15} \sum_{n=0}^{\infty} f^{(n,2,2)}
\Big]\,,
\\
T^{yy} =\;& 
\frac{4\sqrt{\pi}}{(2\pi)^3} \Lambda^4 \Big[
f^{(0,0,0)}-f^{(1,0,0)}
\condbreak{-}
4\sqrt{5} \sum_{n=0}^{\infty} f^{(n,2,0)}
-4\sqrt{15} \sum_{n=0}^{\infty} f^{(n,2,2)}
\Big]\,,
\\
T^{zz} =\;& 
\frac{4\sqrt{\pi}}{(2\pi)^3} \Lambda^4 \Big[
f^{(0,0,0)}-f^{(1,0,0)}
\condbreak{+}
8\sqrt{5} \sum_{n=0}^{\infty} f^{(n,2,0)}
\Big]\,,
\\
T^{xy} =\;& 
\frac{16\sqrt{15\pi}}{(2\pi)^3} \Lambda^4 \Big[
\sum_{n=0}^{\infty} f^{(n,2,-2)}
\Big]\,,
\\
T^{xz} =\;& 
\frac{16\sqrt{15\pi}}{(2\pi)^3} \Lambda^4 \Big[
\sum_{n=0}^{\infty} f^{(n,2,1)}
\Big]\,,
\\
T^{yz} =\;& 
\frac{16\sqrt{15\pi}}{(2\pi)^3} \Lambda^4 \Big[
\sum_{n=0}^{\infty} f^{(n,2,-1)}
\Big]\,.
\label{eq:conservationAndCoefficient_5}
\end{align}

We analyze the thermalization and linearization timescales of the energy-momentum tensor components $T^{\mu\nu}$ to systematically investigate the applicability of hydrodynamics. The results are presented in Figure~\ref{fig:NxLx_Tmunu}, with a graphical setup similar to Figure~\ref{fig:NxLx_Moments}. Panel (a) displays the ensemble-averaged thermalization of $T^{\mu\nu}$, while panel~(b) illustrates the ensemble-averaged linearization process. In both panels, the temporal evolution of each component is fitted to an exponential decay, $\exp[-t/(a\tau)]$; the fastest and slowest fits are highlighted with orange dashed lines. Panel~(c) reports the fit parameters for every component.

\begin{condfigure}
    \centering
    \includegraphics[width=1.0\textwidth]{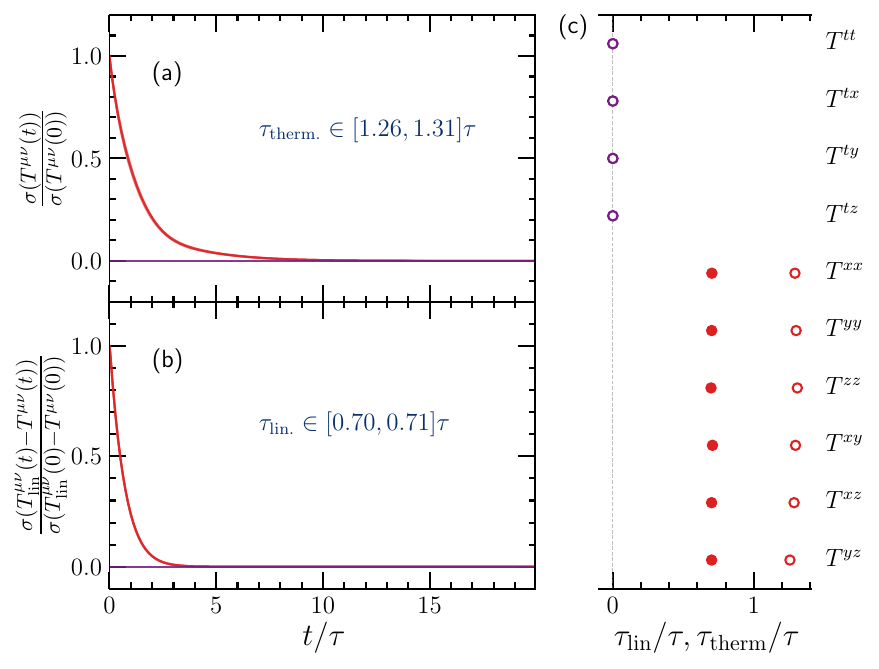}
    \caption{
    Similar to Figure~\ref{fig:NxLx_Moments}, but showing the thermalization (a) and linearization (b) timescales for the energy–momentum tensor $T^{\mu\nu}$ [Eq.~\eqref{eq:NxLx_Tmunu}]. Purple markers denote conserved components; red markers denote nonconserved components. Times are in units of $\tau$.
    }
    \label{fig:NxLx_Tmunu}
\end{condfigure}
The thermalization and linearization statistics in Figure~\ref{fig:NxLx_Tmunu} are consistent with those in Figures~\ref{fig:LB_vs_B}, \ref{fig:N0L6_therm_and_lin}, and \ref{fig:NxLx_Moments}, indicating a robust equilibration pattern. As expected in the homogeneous setup, the conserved quantities---energy density $T^{tt}$ and momentum densities $T^{tx}, T^{ty}, T^{tz}$---exhibit zero thermalization and linearization timescales. In contrast, the six nonconserved components ($T^{xx}, T^{yy}, T^{zz}, T^{xy}, T^{yz}, T^{xz}$) exhibit nearly identical, nonzero thermalization and linearization times. Notably, the linearization time is roughly half the thermalization time, as evidenced by the ratio $0.8/1.5 \approx 0.5$. This indicates that the system reaches a hydrodynamic (linear Boltzmann) regime before achieving full thermal equilibrium.

Collectively, our analysis of both generalized moments and the energy–momentum tensor establishes a robust, observable-insensitive separation between the onset of linearization (hydrodynamic) and full thermalization. Across all moment indices $i, j, k, s$ and for every tensor component, the extracted linearization timescale $\tau_{\mathrm{lin}}$ is consistently about one half of the corresponding thermalization timescale $\tau_{\mathrm{therm}}$. These findings align with the isotropic $(\ell_{\max}=0)$ and angular-only $(n_{\max}=0)$ cases, indicating the robustness of the $\tau_{\mathrm{lin}}/\tau_{\mathrm{therm}}\approx 1/2$ ordering. These results provide new insight into the longstanding early-thermalization puzzle in heavy-ion collisions, suggesting a clear hierarchy between the two timescales.
\section{Conclusion and Discussion}\label{sec:Summary}

In this work, we employ the spectral nonlinear Boltzmann equation---the lowest-order truncation of the spectral BBGKY hierarchy introduced in Ref.~\cite{Lu:2025yry}---to quantitatively compute linearization and thermalization timescales in heavy-ion collisions. The analyses are performed for a homogeneous, massless system with a constant differential cross section [Eq.~\eqref{eq:example_collision_kernel}] across three complementary truncations isolating radial-only, axisymmetric angular-only, and fully coupled dynamics: (i) $\ell_{\mathrm{max}}=0$, a truncation to an isotropic momentum distribution, where all angular dependence is thermalized and only the radial sector remains dynamical; (ii) $n_{\mathrm{max}}=0$ and $m=0$, a truncation to the lowest radial mode, where the radial dependence is fixed to its equilibrium form and only angular modes evolve dynamically; (iii) $\ell_{\mathrm{max}}=4, n_{\mathrm{max}}=4$, the most general case considered here, in which both radial and angular degrees of freedom are retained and evolve away from equilibrium. The first two truncations admit an analytic treatment: although explicit solutions for the distribution functions are not available, the recursive structure of the spectral Boltzmann equations [Eq.~\eqref{eq:spectral_boltzmann_isotropic} for case (i) and Eq.~\eqref{eq:spectral_boltzmann_anisotropic} for case (ii)], together with the bounded dependence of the moments on the expansion coefficients [Eq.~\eqref{eq:energy_moment_isotropic} for case (i) and Eq.~\eqref{eq:anisotropic_moment} for case (ii)], enables precise computation of the moments for arbitrary initial conditions. The third truncation requires numerical analysis, extending the discussion to generic initial states. 

Across all three truncation schemes, we observe a robust ordering: $\tau_{\mathrm{lin}} \approx \tau_{\mathrm{therm}}/2$. This implies that, whether the early-time dynamics are strongly nonlinear or already near equilibrium, the system inevitably enters a linear regime governed by the linearized Boltzmann equation. While the absolute timescales $\tau_{\mathrm{lin}}$ and $\tau_{\mathrm{therm}}$ vary with the truncation, their ratio remains essentially unchanged. At fixed truncation, the extracted times are also insensitive to the choice of distribution-function ensemble, with the ratio even more stable. Likewise, at fixed truncation, the absolute timescales show mild variation across observables (including components of $T^{\mu\nu}$), whereas the ratio remains insensitive. Together, these findings indicate that $\tau_{\mathrm{lin}}/\tau_{\mathrm{therm}} \approx 1/2$ is a robust feature within the present setup.

The earlier onset of linearization---well before full thermalization---provides a quantitative criterion for hydrodynamic applicability. In particular, our results shed new light on the longstanding early-thermalization puzzle: we demonstrate that hydrodynamic applicability begins as soon as the system enters the linear regime, a transition that occurs in roughly half the time required for complete local thermal equilibrium. This finding naturally explains the empirical success of hydrodynamic models at timescales of order $1~\mathrm{fm}/c$~\cite{Shen:2014vra, Nijs:2020roc, Nijs:2020ors}.

Our conclusion also provides a microscopic support to earlier studies of anisotropic hydrodynamics~\cite{Martinez:2010sc, Martinez:2012tu, Ryblewski:2012rr, Bazow:2013ifa, Bazow:2015cha, Tinti:2015xwa, Molnar:2016vvu, McNelis:2021zji, McNelis:2018jho}, which demonstrated that relaxing isotropy assumptions improves the description of early-time dynamics. Within the simplified setup considered here, we demonstrate that hydrodynamic theory can be extended to a system with sizable momentum-space anisotropy.

Our analysis decouples the onset of hydrodynamic behavior from full thermalization, making it unnecessary to assume that thermalization occurs after only a single collision\footnote{In natural units ($c=1$), typical estimates give a relaxation time of about $1~\mathrm{fm}/c$ and a mean free path of about $1~\mathrm{fm}$.}. This provides a concrete framework for modeling far-from-equilibrium dynamics in relativistic heavy-ion collisions. Several limitations of the present study also point to directions for future work. First, our treatment is restricted to spatially homogeneous systems; extending it to spatially inhomogeneous cases will be essential for realistic applications. Second, the implementation assumes an isotropic, constant differential cross section (specified in the two-body center-of-momentum frame). Incorporating more realistic, QCD-motivated kernels is an important next step. Finally, the framework can be generalized to include inelastic processes such as $gg \leftrightarrow ggg$, enabling direct comparison with earlier work involving inelastic channels~\cite{Xu:2014ega}.
\appendix
\acknowledgments

The authors are grateful to colleagues for their support and insightful feedback of this two-year research endeavor.
This work is supported by Tsinghua University under grant Nos. 04200500123, 531205006, 533305009.
We also acknowledge the support by center of high performance computing, Tsinghua University.

\bibliographystyle{JHEP}
\bibliography{biblio.bib}

\providecommand{\href}[2]{#2}\begingroup\raggedright\begin{thebibliography}{10}

\bibitem{Lee:1974kn}
T.D.~Lee, \emph{{Abnormal Nuclear States and Vacuum Excitations}}, \href{https://doi.org/10.1103/RevModPhys.47.267}{\emph{Rev. Mod. Phys.} {\bfseries 47} (1975) 267}.

\bibitem{Shifman:1978bx}
M.A.~Shifman, A.I.~Vainshtein and V.I.~Zakharov, \emph{{QCD and Resonance Physics. Theoretical Foundations}}, \href{https://doi.org/10.1016/0550-3213(79)90022-1}{\emph{Nucl. Phys. B} {\bfseries 147} (1979) 385}.

\bibitem{Shifman:1978by}
M.A.~Shifman, A.I.~Vainshtein and V.I.~Zakharov, \emph{{QCD and Resonance Physics: Applications}}, \href{https://doi.org/10.1016/0550-3213(79)90023-3}{\emph{Nucl. Phys. B} {\bfseries 147} (1979) 448}.

\bibitem{Colangelo:2000dp}
P.~Colangelo and A.~Khodjamirian, \emph{{QCD sum rules, a modern perspective}},  \href{https://arxiv.org/abs/hep-ph/0010175}{{\ttfamily hep-ph/0010175}}.

\bibitem{Shuryak:2003xe}
E.~Shuryak, \emph{{Why does the quark gluon plasma at RHIC behave as a nearly ideal fluid?}}, \href{https://doi.org/10.1016/j.ppnp.2004.02.025}{\emph{Prog. Part. Nucl. Phys.} {\bfseries 53} (2004) 273} [\href{https://arxiv.org/abs/hep-ph/0312227}{{\ttfamily hep-ph/0312227}}].

\bibitem{Shen:2014vra}
C.~Shen, Z.~Qiu, H.~Song, J.~Bernhard, S.~Bass and U.~Heinz, \emph{{The iEBE-VISHNU code package for relativistic heavy-ion collisions}}, \href{https://doi.org/10.1016/j.cpc.2015.08.039}{\emph{Comput. Phys. Commun.} {\bfseries 199} (2016) 61} [\href{https://arxiv.org/abs/1409.8164}{{\ttfamily 1409.8164}}].

\bibitem{Schenke:2010rr}
B.~Schenke, S.~Jeon and C.~Gale, \emph{{Elliptic and triangular flow in event-by-event (3+1)D viscous hydrodynamics}}, \href{https://doi.org/10.1103/PhysRevLett.106.042301}{\emph{Phys. Rev. Lett.} {\bfseries 106} (2011) 042301} [\href{https://arxiv.org/abs/1009.3244}{{\ttfamily 1009.3244}}].

\bibitem{Karpenko:2013wva}
I.~Karpenko, P.~Huovinen and M.~Bleicher, \emph{{A 3+1 dimensional viscous hydrodynamic code for relativistic heavy ion collisions}}, \href{https://doi.org/10.1016/j.cpc.2014.07.010}{\emph{Comput. Phys. Commun.} {\bfseries 185} (2014) 3016} [\href{https://arxiv.org/abs/1312.4160}{{\ttfamily 1312.4160}}].

\bibitem{vanderSchee:2013pia}
W.~van~der Schee, P.~Romatschke and S.~Pratt, \emph{{Fully Dynamical Simulation of Central Nuclear Collisions}}, \href{https://doi.org/10.1103/PhysRevLett.111.222302}{\emph{Phys. Rev. Lett.} {\bfseries 111} (2013) 222302} [\href{https://arxiv.org/abs/1307.2539}{{\ttfamily 1307.2539}}].

\bibitem{Pang:2018zzo}
L.-G.~Pang, H.~Petersen and X.-N.~Wang, \emph{{Pseudorapidity distribution and decorrelation of anisotropic flow within the open-computing-language implementation CLVisc hydrodynamics}}, \href{https://doi.org/10.1103/PhysRevC.97.064918}{\emph{Phys. Rev. C} {\bfseries 97} (2018) 064918} [\href{https://arxiv.org/abs/1802.04449}{{\ttfamily 1802.04449}}].

\bibitem{Du:2019obx}
L.~Du and U.~Heinz, \emph{{(3+1)-dimensional dissipative relativistic fluid dynamics at non-zero net baryon density}}, \href{https://doi.org/10.1016/j.cpc.2019.107090}{\emph{Comput. Phys. Commun.} {\bfseries 251} (2020) 107090} [\href{https://arxiv.org/abs/1906.11181}{{\ttfamily 1906.11181}}].

\bibitem{Heinz:2002un}
U.W.~Heinz and P.F.~Kolb, \emph{{Two RHIC puzzles: Early thermalization and the HBT problem}},  in \emph{{18th Winter Workshop on Nuclear Dynamics}}, 4, 2002 [\href{https://arxiv.org/abs/hep-ph/0204061}{{\ttfamily hep-ph/0204061}}].

\bibitem{Nijs:2020roc}
G.~Nijs, W.~van~der Schee, U.~G{\"u}rsoy and R.~Snellings, \emph{{Bayesian analysis of heavy ion collisions with the heavy ion computational framework Trajectum}}, \href{https://doi.org/10.1103/PhysRevC.103.054909}{\emph{Phys. Rev. C} {\bfseries 103} (2021) 054909} [\href{https://arxiv.org/abs/2010.15134}{{\ttfamily 2010.15134}}].

\bibitem{Nijs:2020ors}
G.~Nijs, W.~van~der Schee, U.~G{\"u}rsoy and R.~Snellings, \emph{{Transverse Momentum Differential Global Analysis of Heavy-Ion Collisions}}, \href{https://doi.org/10.1103/PhysRevLett.126.202301}{\emph{Phys. Rev. Lett.} {\bfseries 126} (2021) 202301} [\href{https://arxiv.org/abs/2010.15130}{{\ttfamily 2010.15130}}].

\bibitem{Wang:1991hta}
X.-N.~Wang and M.~Gyulassy, \emph{{HIJING: A Monte Carlo model for multiple jet production in p p, p A and A A collisions}}, \href{https://doi.org/10.1103/PhysRevD.44.3501}{\emph{Phys. Rev. D} {\bfseries 44} (1991) 3501}.

\bibitem{Gyulassy:1994ew}
M.~Gyulassy and X.-N.~Wang, \emph{{HIJING 1.0: A Monte Carlo program for parton and particle production in high-energy hadronic and nuclear collisions}}, \href{https://doi.org/10.1016/0010-4655(94)90057-4}{\emph{Comput. Phys. Commun.} {\bfseries 83} (1994) 307} [\href{https://arxiv.org/abs/nucl-th/9502021}{{\ttfamily nucl-th/9502021}}].

\bibitem{Zhou:2024ysb}
F.~Zhou, J.~Brewer and A.~Mazeliauskas, \emph{{Minijet quenching in non-equilibrium quark-gluon plasma}}, \href{https://doi.org/10.1007/JHEP06(2024)214}{\emph{JHEP} {\bfseries 06} (2024) 214} [\href{https://arxiv.org/abs/2402.09298}{{\ttfamily 2402.09298}}].

\bibitem{Schwinger:1951nm}
J.S.~Schwinger, \emph{{On gauge invariance and vacuum polarization}}, \href{https://doi.org/10.1103/PhysRev.82.664}{\emph{Phys. Rev.} {\bfseries 82} (1951) 664}.

\bibitem{Casher:1978wy}
A.~Casher, H.~Neuberger and S.~Nussinov, \emph{{Chromoelectric Flux Tube Model of Particle Production}}, \href{https://doi.org/10.1103/PhysRevD.20.179}{\emph{Phys. Rev. D} {\bfseries 20} (1979) 179}.

\bibitem{1983AnPhy.145..340A}
J.~{Ambj{\o}rn} and R.J.~{Hughes}, \emph{{Canonical quantisation in non-Abelian background fields}}, \href{https://doi.org/10.1016/0003-4916(83)90187-2}{\emph{Annals of Physics} {\bfseries 145} (1983) 340}.

\bibitem{Gyulassy:1985oqt}
M.~Gyulassy and A.~Iwazaki, \emph{{QUARK AND GLUON PAIR PRODUCTION IN SU(N) COVARIANT CONSTANT FIELDS}}, \href{https://doi.org/10.1016/0370-2693(85)90711-7}{\emph{Phys. Lett. B} {\bfseries 165} (1985) 157}.

\bibitem{Matsui:1986xp}
T.~Matsui, \emph{{Dynamical Evolution of the Quark - Gluon Plasma and Phenomenology}}, \href{https://doi.org/10.1016/0375-9474(87)90471-4}{\emph{Nucl. Phys. A} {\bfseries 461} (1987) 27C}.

\bibitem{Gelis:2010nm}
F.~Gelis, E.~Iancu, J.~Jalilian-Marian and R.~Venugopalan, \emph{{The Color Glass Condensate}}, \href{https://doi.org/10.1146/annurev.nucl.010909.083629}{\emph{Ann. Rev. Nucl. Part. Sci.} {\bfseries 60} (2010) 463} [\href{https://arxiv.org/abs/1002.0333}{{\ttfamily 1002.0333}}].

\bibitem{Kovner:1995ts}
A.~Kovner, L.D.~McLerran and H.~Weigert, \emph{{Gluon production at high transverse momentum in the McLerran-Venugopalan model of nuclear structure functions}}, \href{https://doi.org/10.1103/PhysRevD.52.3809}{\emph{Phys. Rev. D} {\bfseries 52} (1995) 3809} [\href{https://arxiv.org/abs/hep-ph/9505320}{{\ttfamily hep-ph/9505320}}].

\bibitem{Kovner:1995ja}
A.~Kovner, L.D.~McLerran and H.~Weigert, \emph{{Gluon production from nonAbelian Weizsacker-Williams fields in nucleus-nucleus collisions}}, \href{https://doi.org/10.1103/PhysRevD.52.6231}{\emph{Phys. Rev. D} {\bfseries 52} (1995) 6231} [\href{https://arxiv.org/abs/hep-ph/9502289}{{\ttfamily hep-ph/9502289}}].

\bibitem{Krasnitz:1999wc}
A.~Krasnitz and R.~Venugopalan, \emph{{The Initial energy density of gluons produced in very high-energy nuclear collisions}}, \href{https://doi.org/10.1103/PhysRevLett.84.4309}{\emph{Phys. Rev. Lett.} {\bfseries 84} (2000) 4309} [\href{https://arxiv.org/abs/hep-ph/9909203}{{\ttfamily hep-ph/9909203}}].

\bibitem{Krasnitz:2000gz}
A.~Krasnitz and R.~Venugopalan, \emph{{The Initial gluon multiplicity in heavy ion collisions}}, \href{https://doi.org/10.1103/PhysRevLett.86.1717}{\emph{Phys. Rev. Lett.} {\bfseries 86} (2001) 1717} [\href{https://arxiv.org/abs/hep-ph/0007108}{{\ttfamily hep-ph/0007108}}].

\bibitem{Schenke:2012wb}
B.~Schenke, P.~Tribedy and R.~Venugopalan, \emph{{Fluctuating Glasma initial conditions and flow in heavy ion collisions}}, \href{https://doi.org/10.1103/PhysRevLett.108.252301}{\emph{Phys. Rev. Lett.} {\bfseries 108} (2012) 252301} [\href{https://arxiv.org/abs/1202.6646}{{\ttfamily 1202.6646}}].

\bibitem{bogoliubov1946kineticEnglish}
N.N.~Bogoliubov, \emph{Kinetic equations}, {\emph{Journal of Physics} {\bfseries 10} (1946) 265}.

\bibitem{bogoliubov1946kineticRussian}
N.N.~Bogoliubov, \emph{Kinetic equations}, {\emph{Journal of Experimental and Theoretical Physics} {\bfseries 16} (1946) 691}.

\bibitem{bogoliubov1947kinetic}
N.N.~Bogoliubov and K.P.~Gurov, \emph{Kinetic equations in quantum mechanics}, {\emph{Journal of Experimental and Theoretical Physics} {\bfseries 17} (1947) 614}.

\bibitem{1946JChPh..14..180K}
J.G.~{Kirkwood}, \emph{{The Statistical Mechanical Theory of Transport Processes I. General Theory}}, \href{https://doi.org/10.1063/1.1724117}{\emph{Journal of Chemical Physics} {\bfseries 14} (1946) 180}.

\bibitem{1947JChPh..15...72K}
J.G.~{Kirkwood}, \emph{{The Statistical Mechanical Theory of Transport Processes II. Transport in Gases}}, \href{https://doi.org/10.1063/1.1746292}{\emph{Journal of Chemical Physics} {\bfseries 15} (1947) 72}.

\bibitem{1946RSPSA.188...10B}
M.~{Born} and H.S.~{Green}, \emph{{A General Kinetic Theory of Liquids. I. The Molecular Distribution Functions}}, \href{https://doi.org/10.1098/rspa.1946.0093}{\emph{Proceedings of the Royal Society of London Series A} {\bfseries 188} (1946) 10}.

\bibitem{yvon1935theorie}
J.~Yvon, \emph{La th{\'e}orie statistique des fluides et l'{\'e}quation d'{\'e}tat}, no.~203 in Actualit{\'e}s Scientifiques et Industrielles, Hermann, Paris (1935).

\bibitem{Lu:2025yry}
S.~Lu and S.~Shi, \emph{{Spectral BBGKY: a scalable scheme for nonlinear Boltzmann and correlation kinetics}},  \href{https://arxiv.org/abs/2507.14243}{{\ttfamily 2507.14243}}.

\bibitem{Martinez:2010sc}
M.~Martinez and M.~Strickland, \emph{{Dissipative Dynamics of Highly Anisotropic Systems}}, \href{https://doi.org/10.1016/j.nuclphysa.2010.08.011}{\emph{Nucl. Phys. A} {\bfseries 848} (2010) 183} [\href{https://arxiv.org/abs/1007.0889}{{\ttfamily 1007.0889}}].

\bibitem{Martinez:2012tu}
M.~Martinez, R.~Ryblewski and M.~Strickland, \emph{{Boost-Invariant (2+1)-dimensional Anisotropic Hydrodynamics}}, \href{https://doi.org/10.1103/PhysRevC.85.064913}{\emph{Phys. Rev. C} {\bfseries 85} (2012) 064913} [\href{https://arxiv.org/abs/1204.1473}{{\ttfamily 1204.1473}}].

\bibitem{Ryblewski:2012rr}
R.~Ryblewski and W.~Florkowski, \emph{{Highly-anisotropic hydrodynamics in 3+1 space-time dimensions}}, \href{https://doi.org/10.1103/PhysRevC.85.064901}{\emph{Phys. Rev. C} {\bfseries 85} (2012) 064901} [\href{https://arxiv.org/abs/1204.2624}{{\ttfamily 1204.2624}}].

\bibitem{Bazow:2013ifa}
D.~Bazow, U.W.~Heinz and M.~Strickland, \emph{{Second-order (2+1)-dimensional anisotropic hydrodynamics}}, \href{https://doi.org/10.1103/PhysRevC.90.054910}{\emph{Phys. Rev. C} {\bfseries 90} (2014) 054910} [\href{https://arxiv.org/abs/1311.6720}{{\ttfamily 1311.6720}}].

\bibitem{Bazow:2015cha}
D.~Bazow, U.W.~Heinz and M.~Martinez, \emph{{Nonconformal viscous anisotropic hydrodynamics}}, \href{https://doi.org/10.1103/PhysRevC.91.064903}{\emph{Phys. Rev. C} {\bfseries 91} (2015) 064903} [\href{https://arxiv.org/abs/1503.07443}{{\ttfamily 1503.07443}}].

\bibitem{Tinti:2015xwa}
L.~Tinti, \emph{{Anisotropic matching principle for the hydrodynamic expansion}}, \href{https://doi.org/10.1103/PhysRevC.94.044902}{\emph{Phys. Rev. C} {\bfseries 94} (2016) 044902} [\href{https://arxiv.org/abs/1506.07164}{{\ttfamily 1506.07164}}].

\bibitem{Molnar:2016vvu}
E.~Molnar, H.~Niemi and D.H.~Rischke, \emph{{Derivation of anisotropic dissipative fluid dynamics from the Boltzmann equation}}, \href{https://doi.org/10.1103/PhysRevD.93.114025}{\emph{Phys. Rev. D} {\bfseries 93} (2016) 114025} [\href{https://arxiv.org/abs/1602.00573}{{\ttfamily 1602.00573}}].

\bibitem{McNelis:2021zji}
M.~McNelis, D.~Bazow and U.~Heinz, \emph{{Anisotropic fluid dynamical simulations of heavy-ion collisions}}, \href{https://doi.org/10.1016/j.cpc.2021.108077}{\emph{Comput. Phys. Commun.} {\bfseries 267} (2021) 108077} [\href{https://arxiv.org/abs/2101.02827}{{\ttfamily 2101.02827}}].

\bibitem{McNelis:2018jho}
M.~McNelis, D.~Bazow and U.~Heinz, \emph{{(3+1)-dimensional anisotropic fluid dynamics with a lattice QCD equation of state}}, \href{https://doi.org/10.1103/PhysRevC.97.054912}{\emph{Phys. Rev. C} {\bfseries 97} (2018) 054912} [\href{https://arxiv.org/abs/1803.01810}{{\ttfamily 1803.01810}}].

\bibitem{Xu:2004mz}
Z.~Xu and C.~Greiner, \emph{{Thermalization of gluons in ultrarelativistic heavy ion collisions by including three-body interactions in a parton cascade}}, \href{https://doi.org/10.1103/PhysRevC.71.064901}{\emph{Phys. Rev. C} {\bfseries 71} (2005) 064901} [\href{https://arxiv.org/abs/hep-ph/0406278}{{\ttfamily hep-ph/0406278}}].

\bibitem{Xu:2007jv}
Z.~Xu, C.~Greiner and H.~Stocker, \emph{{PQCD calculations of elliptic flow and shear viscosity at RHIC}}, \href{https://doi.org/10.1103/PhysRevLett.101.082302}{\emph{Phys. Rev. Lett.} {\bfseries 101} (2008) 082302} [\href{https://arxiv.org/abs/0711.0961}{{\ttfamily 0711.0961}}].

\bibitem{Xu:2008av}
Z.~Xu and C.~Greiner, \emph{{Elliptic flow of gluon matter in ultrarelativistic heavy-ion collisions}}, \href{https://doi.org/10.1103/PhysRevC.79.014904}{\emph{Phys. Rev. C} {\bfseries 79} (2009) 014904} [\href{https://arxiv.org/abs/0811.2940}{{\ttfamily 0811.2940}}].

\bibitem{Xu:2007ns}
Z.~Xu and C.~Greiner, \emph{{Shear viscosity in a gluon gas}}, \href{https://doi.org/10.1103/PhysRevLett.100.172301}{\emph{Phys. Rev. Lett.} {\bfseries 100} (2008) 172301} [\href{https://arxiv.org/abs/0710.5719}{{\ttfamily 0710.5719}}].

\bibitem{Kovtun:2004de}
P.~Kovtun, D.T.~Son and A.O.~Starinets, \emph{{Viscosity in strongly interacting quantum field theories from black hole physics}}, \href{https://doi.org/10.1103/PhysRevLett.94.111601}{\emph{Phys. Rev. Lett.} {\bfseries 94} (2005) 111601} [\href{https://arxiv.org/abs/hep-th/0405231}{{\ttfamily hep-th/0405231}}].

\bibitem{Blaizot:2011xf}
J.-P.~Blaizot, F.~Gelis, J.-F.~Liao, L.~McLerran and R.~Venugopalan, \emph{{Bose--Einstein Condensation and Thermalization of the Quark Gluon Plasma}}, \href{https://doi.org/10.1016/j.nuclphysa.2011.10.005}{\emph{Nucl. Phys. A} {\bfseries 873} (2012) 68} [\href{https://arxiv.org/abs/1107.5296}{{\ttfamily 1107.5296}}].

\bibitem{Blaizot:2013lga}
J.-P.~Blaizot, J.~Liao and L.~McLerran, \emph{{Gluon Transport Equation in the Small Angle Approximation and the Onset of Bose-Einstein Condensation}}, \href{https://doi.org/10.1016/j.nuclphysa.2013.10.010}{\emph{Nucl. Phys. A} {\bfseries 920} (2013) 58} [\href{https://arxiv.org/abs/1305.2119}{{\ttfamily 1305.2119}}].

\bibitem{Xu:2014ega}
Z.~Xu, K.~Zhou, P.~Zhuang and C.~Greiner, \emph{{Thermalization of gluons with Bose-Einstein condensation}}, \href{https://doi.org/10.1103/PhysRevLett.114.182301}{\emph{Phys. Rev. Lett.} {\bfseries 114} (2015) 182301} [\href{https://arxiv.org/abs/1410.5616}{{\ttfamily 1410.5616}}].

\bibitem{Mendoza:2009gm}
M.~Mendoza, B.~Boghosian, H.J.~Herrmann and S.~Succi, \emph{{Fast Lattice Boltzmann Solver for Relativistic Hydrodynamics}}, \href{https://doi.org/10.1103/PhysRevLett.105.014502}{\emph{Phys. Rev. Lett.} {\bfseries 105} (2010) 014502} [\href{https://arxiv.org/abs/0912.2913}{{\ttfamily 0912.2913}}].

\bibitem{Romatschke:2011hm}
P.~Romatschke, M.~Mendoza and S.~Succi, \emph{{A fully relativistic lattice Boltzmann algorithm}}, \href{https://doi.org/10.1103/PhysRevC.84.034903}{\emph{Phys. Rev. C} {\bfseries 84} (2011) 034903} [\href{https://arxiv.org/abs/1106.1093}{{\ttfamily 1106.1093}}].

\bibitem{Press2007}
W.H.~Press, S.A.~Teukolsky, W.T.~Vetterling and B.P.~Flannery, \emph{Numerical Recipes 3rd Edition: The Art of Scientific Computing}, Cambridge University Press, USA, 3~ed. (2007).

\bibitem{Bazow:2015dha}
D.~Bazow, G.S.~Denicol, U.~Heinz, M.~Martinez and J.~Noronha, \emph{{Analytic solution of the Boltzmann equation in an expanding system}}, \href{https://doi.org/10.1103/PhysRevLett.116.022301}{\emph{Phys. Rev. Lett.} {\bfseries 116} (2016) 022301} [\href{https://arxiv.org/abs/1507.07834}{{\ttfamily 1507.07834}}].

\bibitem{Strickland:2018ayk}
M.~Strickland, \emph{{The non-equilibrium attractor for kinetic theory in relaxation time approximation}}, \href{https://doi.org/10.1007/JHEP12(2018)128}{\emph{JHEP} {\bfseries 12} (2018) 128} [\href{https://arxiv.org/abs/1809.01200}{{\ttfamily 1809.01200}}].

\end{thebibliography}\endgroup

\end{document}